\documentclass[12pt]{iopart}
\usepackage{iopams}
\usepackage{graphicx}
\usepackage{color}
\usepackage{ulem}
\usepackage{multirow}

\eqnobysec

\begin{document}

\title{Emergent spin-valley-orbital physics by spontaneous parity breaking}

\author{Satoru Hayami$^{1}$, Hiroaki Kusunose$^2$, and Yukitoshi Motome$^3$}

\address{$^1$ Department of Physics, Hokkaido University, Sapporo 060-0810, Japan}
\address{$^2$ Department of Physics, Meiji University, Kawasaki 214-8571, Japan}
\address{$^3$ Department of Applied Physics, University of Tokyo, Tokyo 113-8656, Japan}
\ead{hayami@phys.sci.hokudai.ac.jp}
\vspace{10pt}

\begin{abstract}
The spin-orbit coupling in the absence of spatial inversion symmetry plays an important role in realizing intriguing electronic states in solids, such as topological insulators and unconventional superconductivity. 
Usually, the inversion symmetry breaking is inherent in the lattice structures, and hence, it is not easy to control these interesting properties by external parameters. 
We here theoretically investigate the possibility of generating the spin-orbital entanglement  by spontaneous electronic ordering caused by electron correlations. 
In particular, we focus on the centrosymmetric lattices with local asymmetry at the lattice sites, e.g., zig-zag, honeycomb, and diamond structures. 
In such systems, conventional staggered orders, such as charge order and antiferromagnetic order, break the inversion symmetry and activate the antisymmetric spin-orbit coupling, which is hidden in a sublattice-dependent form in the paramagnetic state. 
Considering a minimal two-orbital model on a honeycomb lattice, we scrutinize the explicit form of the antisymmetric spin-orbit coupling for all the possible staggered charge, spin, orbital, and spin-orbital orders. 
We show that the complete table is useful for understanding of spin-valley-orbital physics, such as spin and valley splitting in the electronic band structure and generalized magnetoelectric responses in not only spin but also orbital and spin-orbital channels, reflecting in peculiar magnetic, elastic, and optical properties in solids. 
\end{abstract}

%
\vspace{2pc}
%
\submitto{\JPCM}
%
%
%

\section{Introduction}
\label{sec:Introduction}

The spin-orbit coupling (SOC), which originates from the relativistic motion of electrons, is a source of interesting electronic states in solids. 
In particular, in the systems without spatial inversion symmetry, the SOC acquires an antisymmetric component with respect to the wave vector~\cite{kane1966k,Winkler201012}. 
This is called the antisymmetric SOC (ASOC), typified by the Rashba SOC near a surface and interface~\cite{rashba1960properties,bychkov1984oscillatory,Koga_PhysRevLett.89.046801}, and the Dresselhaus SOC in the cubic systems lacking inversion symmetry~\cite{Dresselhaus_PhysRev.100.580,Dresselhaus_Dresselhaus_Jorio}. 
The ASOC has been extensively studied as it results in intriguing physics, such as Dirac electrons at the surface of topological insulators~\cite{Hasan_RevModPhys.82.3045,qi2011topological}, the spin Hall effect~\cite{hirsch1999spin,sinova2004universal}, multiferroics~\cite{fiebig2005revival,cheong2007multiferroics}, and the noncentrosymmetric superconductivity~\cite{Bauer_Sigrist201201}. 

\begin{figure}[htb!]
\centering
\includegraphics[width=1.0 \hsize]{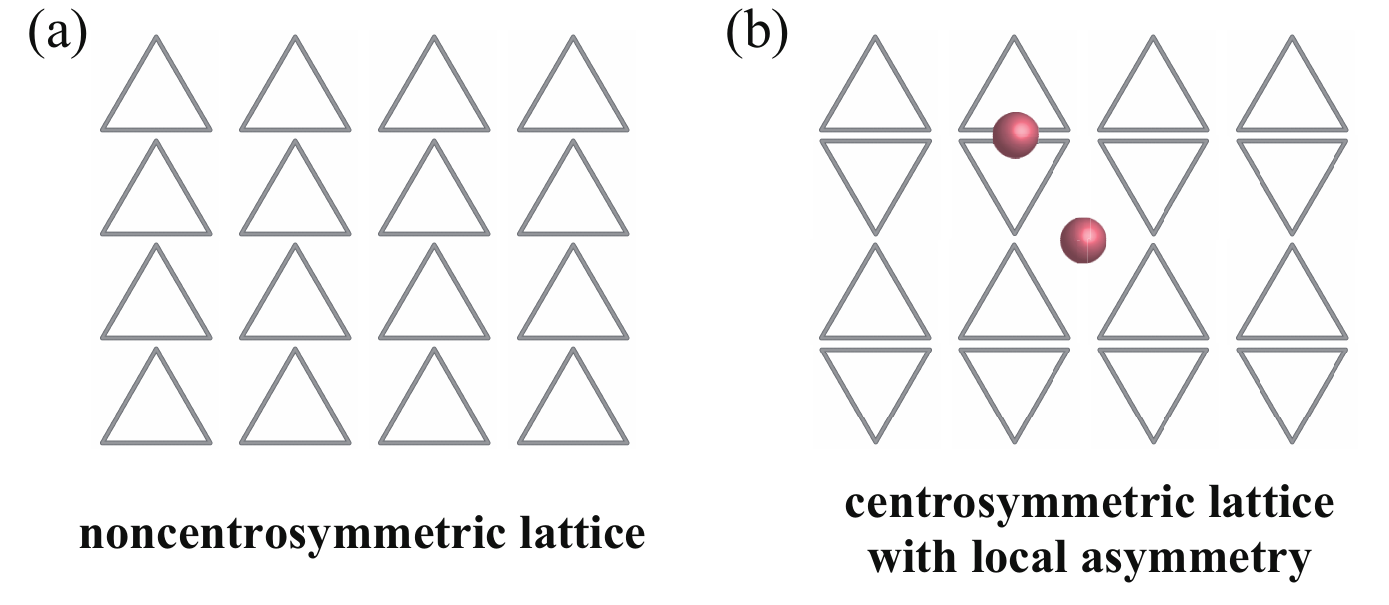} 
\caption{
\label{fig:Lattice_ASOC}
Schematic pictures of (a) a noncentrosymmetric system and (b) a centrosymmetric system with local asymmetry. 
In the latter, spatial inversion symmetry is broken at the lattice sites (vertices on each triangle) despite the presence of inversion symmetry in the whole system. 
The red spheres represent the inversion centers. 
}
\end{figure}

The ASOC is in general described by the simple Hamiltonian in the form of 
\begin{eqnarray}
\mathcal{H}_{{\rm ASOC}} (\bi{k}) = \bi{g}(\bi{k})\cdot \bsigma \sim (\bi{k} \times \bnabla V) \cdot \bsigma, 
\end{eqnarray}
where $\bi{g}(\bi{k})$ is the axial vector, which is asymmetric with respect to $\bi{k}$, $\bsigma$ is the vector of Pauli matrices describing the spin degree of freedom in electrons, and $\bnabla V$ is an asymmetric potential gradient. 
The $\bi{k}$ dependence of $\bi{g}(\bi{k})$ is determined by the potential gradient $\bnabla V$, which  depends on the symmetry of the system. 
The magnitude of ASOC is predominantly determined by three microscopic parameters~\cite{Yanase_kotai,Hayami_PhysRevB.90.024432}: (I) the atomic SOC, (II) transfer integrals between orbitals with different parity, and (III) local hybridizations between orbitals with different parity. 
Note that, while (I) and (II) exist even in the centrosymmetric systems, 
(III) vanishes when the atomic site is an inversion center. 
In order to realize a gigantic ASOC for applications to electronics and spintronics, it is desired to enhance these three parameters. 
However, it is difficult to control them flexibly because such microscopic parameters are intrinsically determined by the lattice structures and the atomic orbitals. 

In the present study, we theoretically explore the possibility of controlling the ASOC by spontaneous symmetry breaking in the electronic degrees of freedom. 
There are two key ingredients, in addition to (I)-(III) above: 
one is a centrosymmetric lattice structure with local asymmetry, and the other is an electronic order which breaks the inversion symmetry spontaneously, as we detail below. 
 
In the first place, we consider a class of centrosymmetric lattices whose inversion centers are located not at the lattice sites but at offsite positions~\cite{zhang2014hidden,Yanase:JPSJ.83.014703,Hayami_PhysRevB.90.024432}. 
A schematic example is shown in figure~\ref{fig:Lattice_ASOC}, in comparison with the noncentrosymmetric case. 
In the figures, the upward and downward triangles, whose vertices represent the lattice sites, are placed periodically in a different manner. 
Figure~\ref{fig:Lattice_ASOC}(a) represents an example of noncentrosymmetric lattice structures; there is no inversion center in the system. 
On the other hand, in figure~\ref{fig:Lattice_ASOC}(b), while the system possesses the inversion symmetry at some intersite positions as displayed by the red spheres, there are no inversion center at every lattice site. 
This type of local asymmetry is found in many lattice structures with the sublattice degree of freedom, for instance, zigzag chain,  honeycomb, and diamond lattices. 
Interestingly, in these lattices, the local asymmetry generates a potential gradient at each lattice site, ${\bnabla V_s }$, which depends on the sublattice $s$. 
This leads to an ASOC in the sublattice-dependent form, whose Hamiltonian is given by 
\begin{eqnarray}
\label{eq:HamASOC_site}
\mathcal{H}_{{\rm ASOC}} (\bi{k}) = \sum_{s} \bi{g}_s (\bi{k}) \cdot \bsigma \sim \sum_s 
(\bi{k} \times \bnabla V_s ) \cdot \bsigma. 
\end{eqnarray}
This is unchanged for the spatial inversion operation with respect to the intersite inversion centers, $\bi{k} \leftrightarrow -\bi{k}$ and $\bnabla V_{s} \leftrightarrow \bnabla V_{s'}=-\bnabla V_{s}$ ($s \neq s'$). 
Therefore, in contrast to the ASOC in noncentrosymmetric systems, this type of sublattice-dependent ASOC does not contribute to bulk properties, such as spin splitting in the band structure and the magnetoelectric effect. 
Thus, equation~(\ref{eq:HamASOC_site}) can be regarded as the ``hidden" ASOC: the system retains ASOC but the net component is zero because of the cancellation between the sublattices. 

The second key ingredient, the spontaneous inversion symmetry breaking by electronic orders, plays a crucial role in activating the ``hidden" ASOC. 
When the electronic order occurs with breaking the inversion symmetry, it induces a net component of ASOC. 
Such an emergent ASOC by spontaneous parity breaking has recently been attracted interest as it brings new aspects in the physics of SOC~\cite{sakhnenko2012magnetoelectric,Yanase:JPSJ.83.014703,Hayami_PhysRevB.90.024432,Hayami_PhysRevB.90.081115,Fu_PhysRevLett.115.026401}. 
This type of ASOC has flexible controllability through the phase transition in the electronic degree of freedom. 
In other words, the ASOC can be varied, or even switched on and off, by external parameters, such as pressure, electric and magnetic fields. 
In addition, the form of the ASOC, i.e., the $\bi{k}$ dependence and amplitude, which is usually determined by the lattice and band structures, is also controllable through the spontaneous electronic orders. 

Furthermore, the emergent ASOC gives rise to intriguing electronic structures and transport properties. 
This is due to the fact that the spontaneous electronic orders in the presence of the ``hidden" ASOC simultaneously accompany multipoles with an odd parity, such as magnetic quadrupole, electric octupole~\cite{hitomi2014electric}, and magnetic toroidal moments~\cite{Yanase:JPSJ.83.014703,Hayami_PhysRevB.90.024432,Hayami_PhysRevB.90.081115,Hayami_1742-6596-592-1-012101,Hayami_doi:10.7566/JPSJ.84.064717,sugita2015antisymmetric}.  
Such odd-parity multipoles bring about a peculiar modulation of the electronic structures and unconventional off-diagonal responses~\cite{volkov1981macroscopic,Spaldin:0953-8984-20-43-434203,kopaev2009toroidal}. 
For example, a staggered antiferromagnetic (AFM) order on the zigzag chain, which accompanies a ferroic toroidal order (see section~\ref{sec:odd-parity}), modifies the electronic band structure in an asymmetric way in the momentum space, and results in unusual off-diagonal responses including the magnetoelectric effect~\cite{Yanase:JPSJ.83.014703,Hayami_PhysRevB.90.024432,Hayami_doi:10.7566/JPSJ.84.064717} and asymmetric modulation of collective spin-wave excitations~\cite{Hayami_doi:10.7566/JPSJ.85.053705}. 
This indicates that the odd-parity multipoles open the further possibility of enriching the spin-orbital entangled phenomena. 

In the previous work~\cite{Hayami_PhysRevB.90.081115}, the authors addressed the issue of emergent ASOC by analyzing a minimal microscopic model on a honeycomb lattice. 
The effects of various charge, spin, orbital, and spin-orbital orders that break spatial inversion symmetry spontaneously were studied by the symmetry analysis as well as the mean-field approximations. 
In this paper, we push forward this issue in a more general framework. 
Specifically, we derive the explicit form of the ASOC for all the possible symmetry breakings including charge, spin, orbital, and spin-orbital channels in the same minimal model. 
We also provide a comprehensive survey of off-diagonal responses as well as the modulations of electronic structures. 
These analyses will clarify how the emergent ASOC gives rise to fascinating properties, such as the spin and valley splitting in the band structure, asymmetric band modulation with a band bottom shift, and peculiar off-diagonal responses, e.g., the spin and valley Hall effects and magnetoelectric effects. 
The results provide a comprehensive reference for further exploration of the ASOC physics, as our method presented here is straightforwardly applicable to other systems with more realistic electronic and lattice structures. 

The rest of the paper is organized as follows. 
In section~\ref{sec:odd-parity}, we describe how the ASOC is hidden in centrosymmetric systems with local asymmetry and how it is activated by conventional electronic orderings. 
We present several examples of odd-parity multipoles induced by the electronic orders, including the magnetic toroidal multipoles. 
In section~\ref{sec:Two-orbital model on a honeycomb lattice}, we introduce a minimal tight-binding model on the honeycomb lattice including both the atomic SOC and electron correlations. 
We present all the possible staggered orders that break spatial inversion symmetry, and categorize them into seven classes from the viewpoint of the symmetry~\cite{Hayami_PhysRevB.90.081115}. 
In section~\ref{sec:Generalized Antisymmetric Coupling}, we investigate the explicit form of the effective ASOC resulting from the spontaneous electronic ordering. 
We provide the complete table of the emergent ASOC for all seven classes. 
Then, we examine the nature of the ordered states in seven classes as well as the paramagnetic state in detail, focusing on the spin and valley splitting in the band structure, band deformation with a band bottom shift, spin and valley Hall effects in section~\ref{sec:Electronic Structure}, and their off-diagonal responses to an electric current which include magnetoelectric effects in section~\ref{sec:Cross Correlation under Antisymmetric Coupling}. 
Finally, section~\ref{sec:Summary} is devoted to summary and perspectives for future study.
In \ref{Sec:Derivation of Effective Antisymmetric Coupling}, we discuss the derivation of effective ASOC.
We also describe the effect of ASOC on the band structure from the viewpoint of the eigenvalue analysis in~\ref{sec:Comparison between the antisymmetric spin-orbit coupling and eigenvalues}. 

\section{Hidden antisymmetric spin-orbit coupling and odd-parity multipoles}
\label{sec:odd-parity}

\begin{figure}[htb!]
\centering
\includegraphics[width=1.0 \hsize]{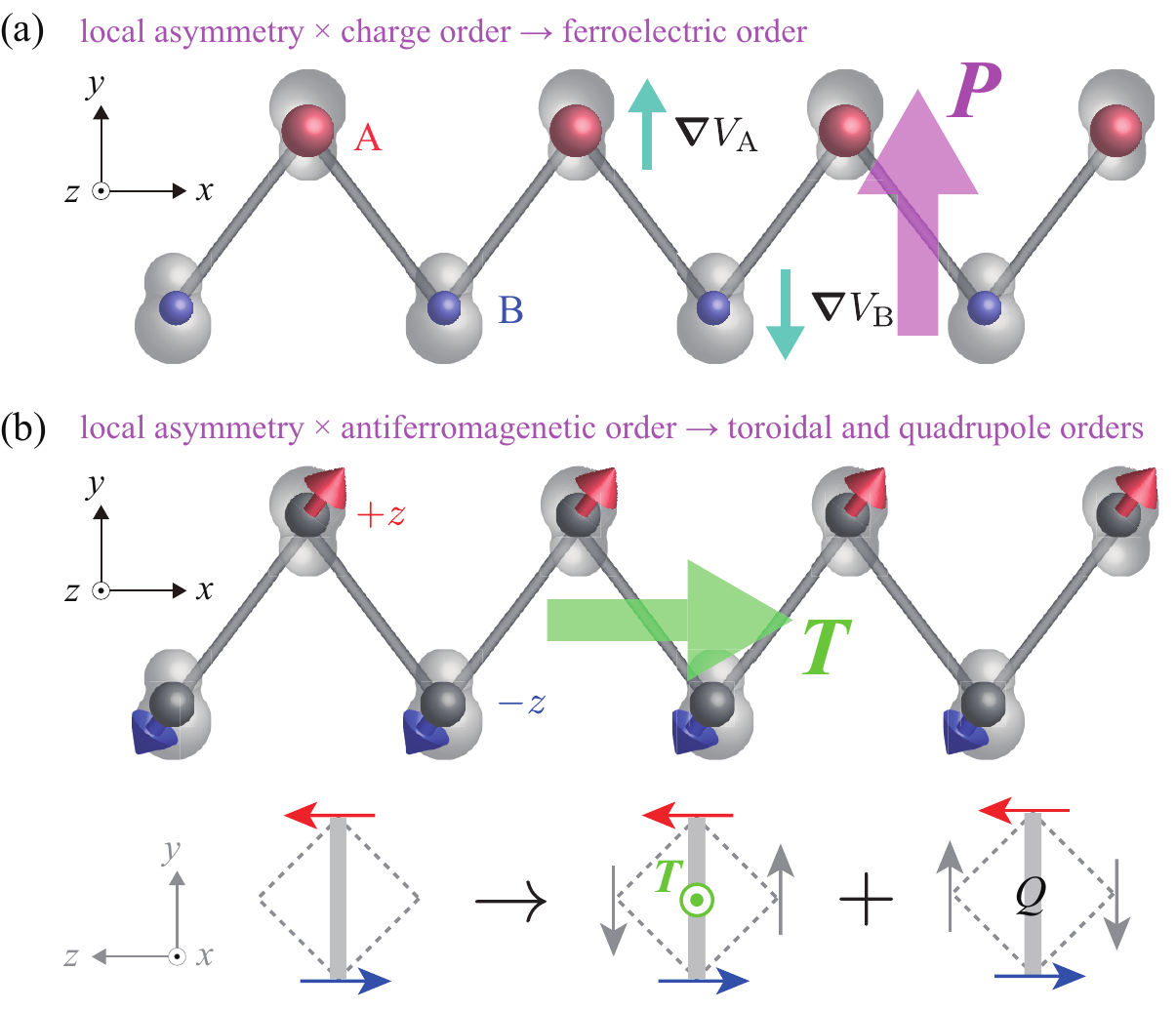} 
\caption{
\label{fig:Zigzag_ASOC}
Schematic pictures for (a) charge ordering on the zigzag chain, which leads to a ferroelectric order (uniform alignment of electric dipoles $\bi{P}$ in the $y$ direction), and (b) antiferromagnetic 
ordering with aligning the magnetic moments along the $z$ direction, which induces a ferroic order of toroidal moments $\bi{T}$ in the $x$ direction and magnetic quadrupoles $Q$; see the bottom panel. 
The red and blue spheres in (a) represent the local charge densities, while the arrows in (b) the local magnetic moments. 
}
\end{figure}

In this section, we describe how odd-parity multipoles are induced by the spontaneous electronic ordering on centrosymmetric lattices with local asymmetry. 
Let us consider an example of the one-dimensional zigzag chain, as shown in figure~\ref{fig:Zigzag_ASOC}. 
By taking the $x$ axis in the chain direction and the $z$ axis in the perpendicular direction to the plane on which the zigzag chain lies, the potential gradient (local electric field) appears in the $y$ direction with alternating sign between the sublattices A and B due to the lack of the inversion symmetry at each lattice site. 
Therefore, taking $\bi{k} \parallel x$ and $\bnabla V_s \parallel y$ in equation~(\ref{eq:HamASOC_site}), we obtain the ASOC with $\bi{g}_s (\bi{k}) \propto (0, 0, k_x \partial V_s /\partial y)$. 
Thus, $\bi{g}_s (\bi{k})$ has the opposite sign for A and B sublattices and it is canceled out in the whole system. 
This is an example of the ``hidden" ASOC mentioned in the previous section. 

In the presence of such hidden ASOC, spontaneous symmetry breaking by electron correlations can activate the odd-parity multipoles, such as magnetic quadrupole, electric dipole, and toroidal dipole. 
For example, when a charge ordering occurs on the zigzag chain as $n_{{\rm A}} \neq n_{{\rm B}}$ [$n_{{\rm A}({\rm B})}$ is the electronic density in the A (B) sublattice], a uniform electric polarization is induced in the $y$ direction, $P_y \propto (n_{{\rm A}} - n_{{\rm B}})|\partial V_{s}/\partial y|$ 
[figure~\ref{fig:Zigzag_ASOC}(a)].  
This is regarded as the ferroelectric ordering resulting from the uniform alignment of electric dipoles. 
Meanwhile, when we consider a staggered collinear magnetic ordering in the $z$ direction given by $m_{{\rm A}}^z =- m_{{\rm B}}^z$ [$m_{{\rm A}(\rm B)}^z$ is the magnetic moment in the $z$ direction in the A (B) sublattice], the magnetic toroidal moment is induced, together with the magnetic quadrupole moment [figure~\ref{fig:Zigzag_ASOC}(b)]. 
Here, the toroidal moment $\bi{t}$ is defined by $\bi{t} \propto \sum_i (\bi{r}_i \times \bi{S}_i)$, where $\bi{r}_i$ is the position vector from the inversion center to the lattice site $i$ and $\bi{S}_i$ is the magnetic moment at site $i$~\cite{Spaldin:0953-8984-20-43-434203,kopaev2009toroidal}. 
By using this expression, we find that the toroidal moment is induced in the $x$ direction: $T_x \propto (m_{{\rm A}}^{z}-m_{{\rm B}}^{z}) |\partial V_{s}/\partial y|$.  

\begin{figure}[htb!]
\centering
\includegraphics[width=0.95 \hsize]{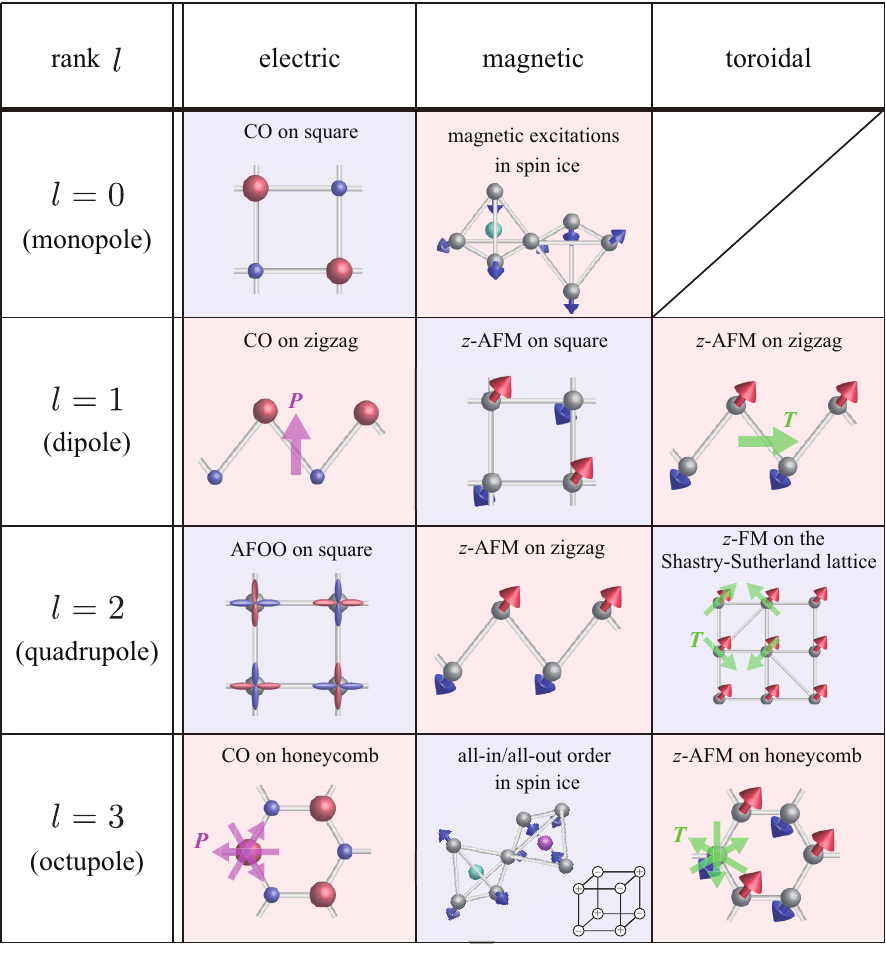} 
\caption{
\label{fig:Table_parity}
Examples of the electric, magnetic, and toroidal magnetic multipoles induced by spontaneous electronic ordering, up to the rank $l=3$. 
The red (blue) boxes stand for multipoles with odd (even) parity. 
CO, FM, AFM, and AFOO represents the charge ordered, ferromagnetic, antiferromagnetic, and antiferroorbital ordered states, respectively. 
The prefix $z$ indicates that the magnetic moments are along the $z$ direction (perpendicular to the lattice plane). 
The size of the spheres reflects the magnitude of local charge densities, and the arrows represent local magnetic moments. 
The spheres in the tetrahedron in the magnetic multipoles with $l=0$ and $l=3$ represent the magnetic monopoles and antimonopoles. 
The toroidal dipoles $\bi{T}$ and electric dipoles $\bi{P}$ are also shown in the figure. 
The cubic in the box of the $l=3$ magnetic multipole represents the schematic arrangement of the magnetic monopoles ($+$) and antimonopoles ($-$) in a staggered way. See the main text for details. 
}
\end{figure}

As easily imagined from the above examples, a variety of multipoles can be realized according to the lattice structures and types of electronic orders. 
We show representative examples in figure~\ref{fig:Table_parity}. 
The second and third columns represent the electric and magnetic multipoles with the rank $l$ in the first column, respectively. 
The last column shows the magnetic toroidal multipoles, which appear in the multipole expansion of a vector potential~\cite{Spaldin:0953-8984-20-43-434203,kopaev2009toroidal,Hayami_PhysRevB.90.024432}. 
 
In the table, the blue boxes represent the multipoles with even parity. 
These are conventional multipoles, which appear even in the presence of spatial inversion symmetry at the lattice sites. 
The well-known examples of the even-parity multipoles are the electric charge (monopole) ($l=0$), magnetic dipole ($l=1$), and electric quadrupole ($l=2$). 
In these examples, the multipoles are defined at each lattice site, and they are aligned in a staggered way, as shown in figure~\ref{fig:Table_parity}. 
In other words, the minimal unit of multipoles is a single lattice site for these cases. 
On the other hand, the multipoles can be defined as spatially extended objects over several lattice sites, as described above. 
The toroidal quadrupole ($l=2$) and magnetic octupole ($l=3$) in figure~\ref{fig:Table_parity} are such examples of even-parity multipoles. 
In the former, the toroidal dipole moments $\bi{T}$ are aligned in the lattice plane to form quadrupoles, as shown in the figure. 
Meanwhile, in the latter, the all-in/all-out ordering of magnetic moments on the pyrochlore lattice is regarded as a magnetic octupole order from the symmetry point of view~\cite{castelnovo2008magnetic,khomskii2012electric,Arima_doi:10.7566/JPSJ.82.013705}. 
We note that this is also interpreted as an antiferroic monopole ordering on the cubic lattice (the unit is a pair of upward and downward tetrahedrons), as shown in the inset~\footnote{In the similar way, the staggered charge ordering on the square lattice (as shown in $l=0$ electric monopole) is regarded as the electric quadrupole ordering in the unit of four sublattices.}. 
 
On the other hand, the red boxes represent rather unconventional odd-parity multipoles, which appear only when the spatial inversion symmetry is broken at the lattice sites. 
It is worth noting that the odd-parity multipoles are realized by conventional charge and magnetic orderings on the centrosymmetric lattices with local asymmetry, except for the magnetic monopole ($l=0$)~\footnote{The magnetic monopole corresponds to a magnetic excitation in spin ice~\cite{bramwell2001spin,castelnovo2008magnetic,khomskii2012electric}.}. 
We have already described the cases in the electric dipole ($l=1$), toroidal dipole ($l=1$), and magnetic quadrupole ($l=2$) by taking the example of the zigzag chain in figure~\ref{fig:Zigzag_ASOC}. 
The odd-parity multipoles on the zigzag chain are naturally extended to those on the honeycomb lattice, since the honeycomb lattice is constructed from the superposition of the zigzag chains connected by threefold rotational symmetry; for instance, the superposition of the toroidal (electric) dipoles connected by threefold rotational symmetry results in the the toroidal (electric) octupole ($l=3$), as shown in figure~\ref{fig:Table_parity}. 

Thus, the spontaneous symmetry breaking by 
conventional electronic ordering can induce unconventional odd-parity multipoles. 
This is unique nature of the centrosymmetric lattice systems with local asymmetry. 
Once these odd-parity multipoles are activated, new quantum states accompanying such as an antisymmetric spin splitting and off-diagonal responses, e.g., a magnetoelectric response, are expected. 
Furthermore, such off-diagonal responses are controllable through the phase transition in the electronic degrees of freedom. 
In the following sections, taking a two-orbital model on the honeycomb lattice as a fundamental example, we examine how the ASOC is generated by spontaneous inversion symmetry breaking, what types of odd-parity multipoles are activated, and how they influence the electronic and transport properties. 

\section{Two-orbital model on a honeycomb lattice}
\label{sec:Two-orbital model on a honeycomb lattice}

In this section, we present a model to investigate spin-orbital coupled phenomena induced by spontaneous electronic ordering. 
In section~\ref{sec:Model Hamiltonian}, we introduce the model Hamiltonian including charge, spin, and orbital degrees of freedom on a honeycomb lattice. 
We show how staggered-type electronic orders break symmetries in the honeycomb-lattice model in section~\ref{sec:Effect of spontaneous parity breaking}. 

\subsection{Model Hamiltonian}
\label{sec:Model Hamiltonian}

\begin{figure}[htb!]
\centering
\includegraphics[width=1.0 \hsize]{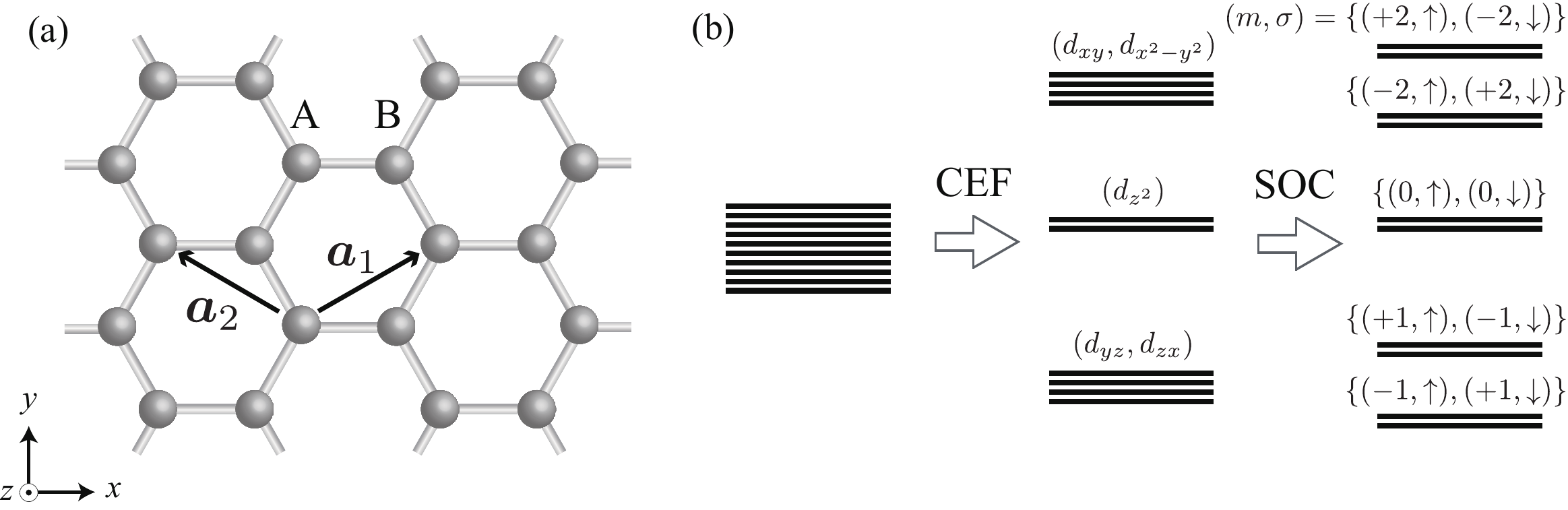} 
\caption{ 
\label{fig:Honeycomb_Lattice}
(a) Schematic picture of a honeycomb lattice; 
the primitive translational vectors are taken as $\bi{a}_1 = (\sqrt{3}/2, 1/2)a$ and $\bi{a}_2 = (-\sqrt{3}/2, 1/2)a$, where the lattice spacing is given by $a/\sqrt{3}$ and we set $a=1$ throughout this paper. 
A and B represent the sublattices. 
(b) Schematic picture of the atomic energy levels of $d$ orbitals. 
We focus on four orbitals $(m, \sigma) = \{ (+1, \uparrow), (-1, \downarrow),(+1, \downarrow), (-1, \uparrow)  \}$ after the level splitting by the trigonal crystalline electric field (CEF) and the atomic SOC. 
}
\end{figure}

For studying the spontaneous breaking of spatial inversion symmetry by electron correlations, we consider a minimal model Hamiltonian on the honeycomb lattice [figure~\ref{fig:Honeycomb_Lattice}(a)]. 
We include the orbital degree of freedom for $d$-electron systems under a crystalline electric field, such as 
the trigonal, trigonal prismatic, and square antiprismatic configuration of the ligands, which splits the atomic energy levels by the magnetic quantum number $m=\pm  2$ ($d_{x^2-y^2}$ and $d_{xy}$ orbitals), $m=\pm 1$ ($d_{zx}$ and $d_{yz}$), and $m=0$ ($d_{z^2}$), as schematically shown in figure~\ref{fig:Honeycomb_Lattice}(b). 
Among them, we take into account only two orbitals with $m=\pm1$~\footnote{The following results are straightforwardly generalized for the $m=\pm2$ case.}.

The Hamiltonian $\mathcal{H}$, which we consider in the present study, consists of the one-body part $\mathcal{H}_0$ and the two-body part $\mathcal{H}_1$: 
\begin{eqnarray}
\label{Eq:Ham}
\mathcal{H}=&\mathcal{H}_0 + \mathcal{H}_1, \\
\label{Eq:H0}
\mathcal{H}_{0} 
=& -t_0 \sum_{\bi{k} m \sigma}
(  \gamma_{0,\bi{k}} c_{{\rm A}\bi{k}m\sigma}^{\dagger} c_{{\rm B}\bi{k} m\sigma} + {\rm H.c.})  -t_1 \sum_{\bi{k}m\sigma}
(\gamma_{m,\bi{k}}  c_{{\rm A}\bi{k}m\sigma}^{\dagger} c_{{\rm B}\bi{k}  \bar{m}\sigma} + {\rm H.c.}) \nonumber
\\ 
&   +\frac{\lambda}{2}\sum_{im\sigma} c_{im\sigma}^{\dagger} (m\sigma) c_{im\sigma}, \\
\label{Eq:H1}
\mathcal{H}_{1} =& \sum_{i\sigma \sigma'}\sum_{m n m' n'}
\frac{U_{m n m' n'}}{2} c_{ i m\sigma}^{\dagger}c_{ i n\sigma'}^{\dagger}
c_{ i n'\sigma'} c_{ i m' \sigma}   + \sum_{\langle i, j \rangle\sigma \sigma'}\sum_{mm'} 
V n_{ i m \sigma}n_{j m' \sigma'}, 
\end{eqnarray}
where $c_{s \bi{k} m\sigma}^{\dagger}$ ($c_{s \bi{k} m\sigma}$) is the creation (annihilation) operator for sublattice $s=$ A or B, wave vector $\bi{k}$, orbital $m=\pm 1$ ($\bar{m}=-m$), and spin $\sigma = \uparrow$ or $\downarrow$; $c_{im\sigma}^\dagger$ ($c_{im\sigma}$) is the real-space representation. 

The first and second terms of $\mathcal{H}_0$ in equation~(\ref{Eq:H0}) represent the intra- and inter-orbital hoppings between nearest-neighbor sites, respectively. 
The sum of $\bi{k}$ is taken over the folded Brillouin zone throughout this paper.
The $\bi{k}$ dependence 
in the hopping terms is given by 
\begin{eqnarray} 
\gamma_{n,\bi{k}}
= \sum_{j=1}^{3}\omega^{(j-1)n}e^{{\rm i}\bi{k}\cdot\bi{\eta}_{j}}
=\gamma_{-n,-\bi{k}}^{*}
\quad
(n=0,\pm1),
\label{eq:gamma}
\end{eqnarray}
where $\omega = e^{2\pi {\rm i}/3}$;
$\bfeta_1 = (\bi{a}_1 - \bi{a}_2)/3$, $\bfeta_2 = (\bi{a}_1 + 2 \bi{a}_2)/3$, and $\bfeta_3 =-(2 \bi{a}_1 + \bi{a}_2)/3$ [$\bi{a}_1$ and $\bi{a}_2$ are primitive translational vectors as shown in figure~\ref{fig:Honeycomb_Lattice}(a)]. 
The additional phase factors $\omega^{\pm n}$ in equation~(\ref{eq:gamma}) come from the angular-momentum transfers between different orbitals, which brings about the asymmetry with respect to $\bi{k}$ and results in peculiar off-diagonal responses in Sec.~\ref{sec:Cross Correlation under Antisymmetric Coupling}. 

The third term in equation~(\ref{Eq:H0}) represents the atomic SOC. 
This term has a nonzero matrix element only for the diagonal component with respect to orbitals (the Ising type) as we consider only $m=\pm 1$. 
The SOC splits the atomic energy levels represented by $(m, \sigma)=(+1,\uparrow)$, $(+1, \downarrow)$, $(-1, \uparrow)$, and $(-1, \downarrow)$ into two groups, $\{ (+1,\uparrow), (-1, \downarrow) \}$ and $\{(+1, \downarrow), (-1, \uparrow)\}$, as shown in the rightmost panel in figure~\ref{fig:Honeycomb_Lattice}(b). 

Meanwhile, the first term of $\mathcal{H}_1$ in equation~(\ref{Eq:H1}) stands for the on-site Coulomb interaction. 
We take $U_{mmmm}=U$, and $U_{mnmn}= U'= U-2J_{{\rm H}}$, and $U_{mnnm} =U_{mmnn}=J_{{\rm H}}$ ($m\neq n$), where $U$, $U'$, and $J_{\rm H}$ are the intra-orbital repulsion, inter-orbital repulsion, and Hund's-rule coupling, respectively. 
The second term in equation~(\ref{Eq:H1}) is the Coulomb repulsion between nearest-neighbor sites, which is introduced to discuss a charge order; here, $n_{im\sigma} = c_{im\sigma}^\dagger c_{im\sigma}$. 

In the following sections, we set $2t_0=1$ as the energy unit. 
We define the electron density as $n_{{\rm e}} = \sum_{s\bi{k} m \sigma} \langle c_{s\bi{k} m \sigma}^{\dagger} c_{s\bi{k} m\sigma} \rangle/(N_s N_{\bi{k}})$, where $N_s$ is the number of sublattices 
and $N_{\bi{k}}$ is the number of grid points in the folded Brillouin zone. 

\subsection{Effect of spontaneous parity breaking}
\label{sec:Effect of spontaneous parity breaking}

We focus on four symmetries in the noninteracting Hamiltonian $\mathcal{H}_0$ in equation~(\ref{Eq:H0})~\cite{Hayami_PhysRevB.90.081115}: spatial inversion ($\mathcal{P}$), time-reversal ($\mathcal{T}$), $2\pi/3$ rotation around the $z$ axis ($\mathcal{R}$), and mirror for the $xz$ plane ($\mathcal{M}$).
These symmetry operations are represented by using three Pauli matrices, $\brho$ for sublattice, $\bsigma$ for spin, and $\btau$ for orbital indices, as 
\begin{eqnarray}
\label{eq:ops}
\mathcal{P}: \rho_x, \qquad 
\mathcal{T}: {\rm i} \sigma_{y}\tau_x K, \qquad 
\mathcal{R}: O e^{2 \pi {\rm {\rm i}} \tau_z /3}, \qquad 
\mathcal{M}: {\rm i} \sigma_{z}, 
\end{eqnarray}
where $K$ is a complex conjugation operator and $O$ is the cyclic permutation operator of the site indices around the rotation center. 
We note that $O$ transforms as $O\gamma_{n,\bi{k}}=\omega^{-n}\gamma_{n,\bi{k}}$. 
The orbital operators, $\tau_{x}$, $\tau_{y}$, and $\tau_{z}$, correspond to the electric quadrupoles, $l_{x}^{2}-l_{y}^{2}$, $l_{x}l_{y}+l_{y}l_{x}$, and the magnetic dipole $l_{z}$, respectively, in the $m=\pm1$ subspace.
Thus, the former two are time-reversal even, and the latter one is time-reversal odd.
The transformation properties of relevant operators and quantities are summarized in table~\ref{tab2}.

\begin{table*}[htb!]
\begin{center}
\caption{
The transformation of relevant operators, matrix elements, and irreducible functions appearing in the effective ASOC in the following sections.
$\mathcal{R}$ transforms the two-dimensional representation like $\bi{r}'=R_{z}(2\pi/3)\,\bi{r}$, where $R_{z}(2\pi/3)$ is $2\pi/3$ rotation matrix around the $z$ axis and two-dimensional vector $\bi{r}=(x,y)$. 
$f_{\rm E}(\bi{k})=f_{{\rm E}_1}(\bi{k})-{\rm i}f_{{\rm E}_2}(\bi{k})$.
}
\scalebox{0.85}{
\begin{tabular}{lccccccc} \hline\hline
        && $\mathcal{P}$ ($\rho_{x}$) & $\mathcal{T}$ (${\rm i}\sigma_{y}\tau_{x}K$) & $\mathcal{R}$ ($O\,e^{2\pi {\rm i}\tau_{z}/3}$) & $\mathcal{M}$ (${\rm i}\sigma_{z}$) && note \\ \hline
$\sigma_{x}$ && $\sigma_{x}$ & $-\sigma_{x}$ & $\sigma_{x}$ & $-\sigma_{x}$ \\
$\sigma_{y}$ && $\sigma_{y}$ & $-\sigma_{y}$ & $\sigma_{y}$ & $-\sigma_{y}$ \\
$\sigma_{z}$ && $\sigma_{z}$ & $-\sigma_{z}$ & $\sigma_{z}$ & $\sigma_{z}$ \\ \hline
$\tau_{x}$ && $\tau_{x}$ & $\tau_{x}$ & $\tau_{x}'$ & $\tau_{x}$ && $l_{x}^{2}-l_{y}^{2}$ \\
$\tau_{y}$ && $\tau_{y}$ & $\tau_{y}$ & $\tau_{y}'$ & $\tau_{y}$ && $l_{x}l_{y}+l_{y}l_{x}$ \\
$\tau_{z}$ && $\tau_{z}$ & $-\tau_{z}$ & $\tau_{z} $ & $\tau_{z}$ && $l_{z}$ \\
$\tau_{\pm}$ && $\tau_{\pm}$ & $\tau_{\mp}$ & $\omega^{\pm1}\tau_{\pm} $ & $\tau_{\pm}$ && $\tau_{\pm}=\tau_{x}\pm {\rm i}\tau_{y}$ \\ \hline
$\rho_{x}$ && $\rho_{x}$ & $\rho_{x}$ & $\rho_{x}$ & $\rho_{x}$ \\
$\rho_{y}$ && $-\rho_{y}$ & $-\rho_{y}$ & $\rho_{y}$ & $\rho_{y}$ \\
$\rho_{z}$ && $-\rho_{z}$ & $\rho_{z}$ & $\rho_{z}$ & $\rho_{z}$ \\
$\rho_{\pm}$ && $\rho_{\mp}$ & $\rho_{\pm}$ & $\rho_{\pm}$ & $\rho_{\pm}$ && $\rho_{\pm}=\rho_{x}\pm {\rm i}\rho_{y}$ \\ \hline\hline
$\gamma_{n,\bi{k}}$ && $\gamma_{-n,\bi{k}}^{*}$ & $\gamma_{-n,\bi{k}}$ & $\omega^{-n}\gamma_{n,\bi{k}}$ & $\gamma_{n,\bi{k}}$ && $\gamma_{n,\bi{k}}=\gamma_{-n,-\bi{k}}^{*}$ \\ \hline
$f_{\rm A}(\bi{k})$ && $-f_{\rm A}(\bi{k})$ & $-f_{\rm A}(\bi{k})$ & $f_{\rm A}(\bi{k})$ & $f_{\rm A}(\bi{k})$ &$\ $& $-\frac{1}{16}k_{y}(3k_{x}^{2}-k_{y}^{2})$ for $\bi{k} \to 0$ \\
$f_{\rm E}(\bi{k})$ &$\qquad$& $-f_{\rm E}(\bi{k})$ & $-f_{\rm E}(\bi{k})$ & $\omega^{-1}f_{\rm E}(\bi{k})$ & $f_{\rm E}(\bi{k})$ && $-\frac{3{\rm i}}{2}(k_{x}+{\rm i}k_{y})$ for $\bi{k} \to 0$ \\
\hline\hline
\end{tabular}
}
\label{tab2}
\end{center}
\end{table*}

These symmetries are spontaneously broken once a phase transition is caused by electron correlations. 
We here consider only the staggered electronic orders with an alternative sign between the A and B sublattices on the bipartite honeycomb lattice, as they are the simplest realizations of the breaking of $\mathcal{P}$ symmetry (spontaneous parity breaking). 
The staggered orders commonly include the component of $\rho_{z}$ with the ordering wave vector $\bi{Q}=(0,0)$ (for simplicity, we omit the orderings represented by $\rho_x$ and $\rho_y$). 
There are sixteen candidates for such staggered orders, which are denoted as 
$\sigma_{\alpha}\tau_{\beta}$ ($\alpha,\beta=0,x,y,z$): a charge order $\sigma_0 \tau_0$ (CO), three spin orders $\sigma_{\mu}\tau_0$ ($\mu$-SO), three orbital orders $\sigma_0 \tau_{\mu}$ ($\nu$-OO), and nine spin-orbital orders $\sigma_{\mu}\tau_{\nu}$ ($\mu\nu$-SOO). 
Here, $\mu, \nu = x,y,z$, and $\sigma_{0}$ and $\tau_0$ are $2\times 2$ unit matrices. 
The symmetry-breaking fields corresponding to these orders are obtained in a general form through the mean-field decoupling of two-body interactions in equation~(\ref{Eq:H1}): 
\begin{eqnarray}
\label{eq: Ham_MF}
\tilde{\mathcal{H}}_{1}  =   -h \sum_{s \bi{k} \sigma \sigma'}
\sum_{m m'} c^{\dagger}_{s  \bi{k} m\sigma} 
\left[p(s) \sigma_{\alpha}\tau_{\beta}
\right]^{\sigma\sigma'}_{mm'}
c_{s \bi{k} m'\sigma'}, 
\end{eqnarray}
where $h$ is the magnitude of the symmetry-breaking fields and $p(s)=+1$ $(-1)$ for $s={\rm A}$ 
(B). 
In the following sections, we deal with symmetry-breaking fields in equation~(\ref{eq: Ham_MF}) instead of the two-body Hamiltonian in equation~(\ref{Eq:H1}). 

\begin{table*}[htb!]
\begin{center}
\caption{
Eight symmetry classes of the paramagnetic state and sixteen staggered ordered states categorized in terms of the presence ($\bigcirc$) or absence ($\times$) of the four symmetries of the system: spatial inversion ($\mathcal{P}$), time-reversal ($\mathcal{T}$), $2\pi/3$ rotation around the $z$ axis ($\mathcal{R}$), and mirror symmetry for the $xz$ plane ($\mathcal{M}$)~\cite{Hayami_PhysRevB.90.081115}. 
PM stands for the paramagnetic state. 
CO, SO, OO, and SOO represent charge, spin, orbital, and spin-orbital orders, respectively, and the prefixes like $z$ denote the type of orders. 
In the columns for SS, VS, and BD, the checkmark ($\checkmark$) shows that the spin splitting, valley splitting, and band deformation with a band bottom shift in the electronic structure takes place under the corresponding order, respectively. 
The columns for sME(u), sME(s), oME(u), and oME(s) indicate the magnetoelectric responses; the prefix s and o represent spin and orbital, respectively, and the u and s in the parentheses denote the uniform and staggered component, respectively. 
See the main text for details. 
}
\scalebox{0.84}{
 \begin{tabular}{cccccccccccccc}
 \hline
 \hline
\#& 
\ \ & $\mathcal{P}$ & $\mathcal{T}$& $\mathcal{R}$& $\mathcal{M}$ &  
& SS & VS & BD & sME(u) & sME(s)& oME(u) & oME(s) \\
\hline
0 &  
PM & $\bigcirc$ & $\bigcirc$ & $\bigcirc$ & $\bigcirc$ & & --  & -- &-- & -- & -- & -- & -- \\ 
1 & CO, $zz$-SOO & $\times$ & $\bigcirc$ & $\bigcirc$ & $\bigcirc$ & & $\checkmark$ & -- & -- & -- & -- & -- & -- \\ 
2 & $x/y$-OO & $\times$ & $\bigcirc$ & $\times$ & $\bigcirc$ & &$\checkmark$ & -- &-- & -- & $\checkmark$ &$\checkmark$ & $\checkmark$ \\ 
3 & $xz/yz$-SOO & $\times$ & $\bigcirc$ & $\bigcirc$ & $\times$ & & $\checkmark$ & -- & -- &  -- & --  & -- & --\\ 
4 & $z$-SO, $z$-OO & $\times$ & $\times$ & $\bigcirc$ & $\bigcirc$ & & -- & $\checkmark$  & -- & -- & --  & -- & --\\ 
5 & $zx/zy$-SOO & $\times$ & $\times$ & $\times$ & $\bigcirc$ & & --  & --& $\checkmark$ & -- & $\checkmark$ & $\checkmark$ & $\checkmark$ \\ 
6 & $x/y$-SO & $\times$ & $\times$ & $\bigcirc$ & $\times$ & & -- & -- & -- & -- & -- & -- & -- \\ 
7 & $xx/yy/xy/yx$-SOO & $\times$ & $\times$ & $\times$ & $\times$ & & -- & -- & -- & $\checkmark$ & $\checkmark$ & -- & $\checkmark$ \\ \hline \hline
\end{tabular}
}
\label{tab}
\end{center}
\end{table*}

The sixteen staggered orders break the symmetries $\mathcal{T}$, $\mathcal{R}$, and $\mathcal{M}$ in a different way. 
We categorize them into seven classes with respect to the symmetries, as summarized in table~\ref{tab}~\cite{Hayami_PhysRevB.90.081115}. 
This symmetry analysis will provide a useful reference for discussing the ASOC induced by each electronic ordering, as described in section~\ref{sec:Generalized Antisymmetric Coupling}.  
It is also helpful for understanding of microscopic and macroscopic physical properties, such as the electronic band structure (section~\ref{sec:Electronic Structure}) and off-diagonal responses including  magnetoelectric effects (section~\ref{sec:Cross Correlation under Antisymmetric Coupling}).  

\section{Antisymmetric Spin-Orbit Coupling Induced by Electronic Ordering}
\label{sec:Generalized Antisymmetric Coupling}
In this section, analyzing the Hamiltonian $\mathcal{H}_0+\tilde{\mathcal{H}_1}$, we show how the effective ASOC is activated by spontaneous symmetry breaking. 
In section~\ref{sec:Paramagnetic state}, we discuss the effective ASOC already existing in the paramagnetic state, which is hidden in the sublattice-dependent form. 
In section~\ref{sec:Ordered states}, we show the explicit form of the ASOC induced by symmetry breaking for several examples of the electronic orders. 
We summarize the ASOCs for all the sixteen possible orders in section~\ref{sec:Summary of Antisymmetric Coupling}. 

\subsection{Paramagnetic state}
\label{sec:Paramagnetic state}

First, we consider the ASOC in the paramagnetic state by taking into account only $\mathcal{H}_0$ in equation~(\ref{Eq:H0}). 
In the paramagnetic state, the ASOC is hidden in the sublattice-dependent form, as discussed in section~\ref{sec:odd-parity}. 
One way to obtain the explicit form of this hidden ASOC is to treat the electron transfers between different sublattices as the perturbation. 
The details of the perturbative calculations are described in~\ref{Sec:Derivation of Effective Antisymmetric Coupling}, and we quote the general result of the effective ASOC in equation~(\ref{eq:effHamASOC}):
\[
H_{\rm eff}(\bi{k})=\sum_{\alpha\beta}[g_{\alpha\beta}^{\rm u}(\bi{k})\rho_{0}+g_{\alpha\beta}^{\rm s}(\bi{k})\rho_{z}]\sigma_{\alpha}\tau_{\beta},
\]
where the term proportional to $\rho_{0}$ ($\rho_{z}$) represents the uniform (staggered) effective ASOC.
As shown in \ref{Sec:Derivation of Effective Antisymmetric Coupling}, 
we find that there are three types of the ASOC in the paramagnetic state, which are represented by 
\begin{eqnarray}
\label{eq:paragz0}
g^{{\rm s}}_{z0} (\bi{k}) \rho_z \sigma_z \tau_0 & \,\sim-\frac{4\sqrt{3}t_{1}^{2}}{\lambda}f_{\rm A}(\bi{k}) 
\rho_z \sigma_z \tau_0, \\
\label{eq:paragzx}
g^{{\rm s}}_{zx} (\bi{k}) \rho_z \sigma_z \tau_x & \,\sim-\frac{4\sqrt{3}t_{0}t_{1}}{\lambda}f_{{\rm E}_1}(\bi{k}) \rho_z \sigma_z \tau_x, \\
\label{eq:paragzy}
g^{{\rm s}}_{zy} (\bi{k}) \rho_z \sigma_z \tau_y & \,\sim-\frac{4\sqrt{3}t_{0}t_{1}}{\lambda}f_{{\rm E}_2}(\bi{k}) \rho_z \sigma_z \tau_y,
\end{eqnarray}
where $f_{\rm A}(\bi{k})$, $f_{{\rm E}_1}(\bi{k})$, and $f_{{\rm E}_2}(\bi{k})$ are given by 
\numparts
\begin{eqnarray}
\label{eq:f_1}
f_{\rm A}(\bi{k}) &= \left[
\cos \left( \frac{\sqrt{3} k_x}{2}\right) - \cos \left(\frac{k_y}{2} \right)
\right]
\sin \left( \frac{k_y}{2} \right),\\
\label{eq:f_2}
f_{{\rm E}_1}(\bi{k}) &= \left[
\cos \left( \frac{\sqrt{3}k_x}{2}\right) + 2 \cos \left( \frac{k_y}{2} \right) 
\right]\sin \left(\frac{k_y}{2}\right), \\
f_{{\rm E}_2}(\bi{k}) &= \sqrt{3} \sin \left( \frac{\sqrt{3}k_x}{2} \right) \cos\left(\frac{k_y}{2} \right), 
\end{eqnarray}
\endnumparts
respectively. 
Note that $f_{\rm A}(\bi{k})$ [$f_{{\rm E}_1}(\bi{k})$ and $f_{{\rm E}_2}(\bi{k})$] and $\tau_{0}$ ($\tau_{x}$ and $\tau_{y}$) belongs to the A (E) irreducible representation of $C_{3}$ group, and $\{f_{{\rm E}_1}(\bi{k})\}^2+\{f_{{\rm E}_2}(\bi{k})\}^2$ has sixfold rotational symmetry. 
Equations~(\ref{eq:paragz0})-(\ref{eq:paragzy}) indicate that these ASOCs are hidden in the sublattice-dependent form and canceled out within the unit cell [$f_{\rm A,E_{1},E_{2}}(\bi{k})\rho_{z}$ is invariant under $\mathcal{P}$]. 

The ASOC in equation~(\ref{eq:paragz0}) has the similar form to the conventional one which has been discussed in topological insulators~\cite{Kane_PhysRevLett.95.146802}. 
The threefold rotational symmetry in the $\bi{k}$ dependence in equation~(\ref{eq:f_1}) reflects the symmetry of the honeycomb lattice. 
In fact, the asymptotic form of $f_{\rm A}(\bi{k})$ in the limit of $\bi{k} \to 0$ is obtained as
\begin{eqnarray}
\label{eq:f1limit}
f_{\rm A}(\bi{k}) \to -\frac{1}{16}k_y (3 k_x^2-k_y^2), 
\end{eqnarray}
which is compatible with the threefold rotational symmetry. 
Owing to the ASOC, the paramagnetic state in our two-orbital Hubbard model shows the quantum spin Hall effect, as discussed in~\cite{Hayami_1742-6596-592-1-012131}. 
The form of ASOC in equation~(\ref{eq:paragz0}) is similar to the effective single-orbital Hubbard model discussed in~\cite{Kane_PhysRevLett.95.146802}. 
However, our two-orbital model exhibits richer behavior in the quantum spin Hall effect than the single-orbital one: our model exhibits several quantized values of the spin Hall conductivity depending on $t_1$ and $\lambda$, at not only 1/2 filling but also 1/4 filling~\cite{Hayami_1742-6596-592-1-012131}. 

On the other hand, equations~(\ref{eq:paragzx}) and (\ref{eq:paragzy}) represent different types of the ASOC: they include not only the spin component but also the orbital component, $\tau_x$ or $\tau_y$. 
As described in section~\ref{sec:Cross Correlation under Antisymmetric Coupling}, they give rise to the coupling between the order parameters and an external electric current. 
The ASOCs in equations~(\ref{eq:paragzx}) and (\ref{eq:paragzy}) have $\bi{k}$-linear contributions in the $\bi{k} \to 0$ limit, as
\begin{equation}
\label{eq:fxylimit}
f_{{\rm E}_1,{\rm E}_2}(\bi{k})\to \frac{3}{2}k_{y,x}. 
\end{equation}
However, this does not mean that these ASOCs break the threefold rotational symmetry because the net ASOC is proportional to the linear combination of the two contributions as $[f_{{\rm E}_1}(\bi{k})\tau_x + f_{{\rm E}_2}(\bi{k})\tau_y]\sigma_{z}\rho_{z}$, which commutes with the threefold rotational operation, $\mathcal{R}$. 
Once the threefold rotational symmetry is broken, $f_{\rm E_{1}}(\bi{k})$ and $f_{\rm E_{2}}(\bi{k})$ become unbalanced, giving rise to linear magnetoelectric couplings, as will be discussed in later sections. 
We summarize the result of the hidden ASOC in the paramagnetic state in table~\ref{tab:ASOC_para} for comparison with the symmetry broken cases discussed in the following sections. 

\begin{table*}[htb!]
\begin{center}
\caption{
Staggered ASOCs, $g^{\rm s}_{\alpha \beta}(\bi{k})$ ($\alpha, \beta=0, x, y, z$), in the paramagnetic state. 
In the table, A, E$_1$, and E$_2$ represent the coefficient of the ASOC as $f_{\rm A}(\bi{k})$, $f_{{\rm E}_1}(\bi{k})$, and $f_{{\rm E}_2}(\bi{k})$, respectively. 
In the presence of threefold rotational symmetry, $f_{{\rm E}_1}$ and $f_{{\rm E}_2}$ appear as the same weight in the ASOC, as they constitute two-dimensional irreducible representation of $C_{3}$ group. 
See also table~\ref{tab:ASOC} for the ordered states. 
}
\label{tab:ASOC_para}
\scalebox{0.9}{
 \begin{tabular}{cccccccccccccccccc}
 \hline
 \hline
\#& 
 & $g^{\rm s}_{00}$ & $g^{\rm s}_{x0}$& $g^{\rm s}_{y0}$& $g^{\rm s}_{z0}$  &$g^{\rm s}_{0x}$ &$g^{\rm s}_{0y}$&$g^{\rm s}_{0z}$&$g^{\rm s}_{xx}$&$g^{\rm s}_{xy}$&$g^{\rm s}_{xz}$&$g^{\rm s}_{yx}$&$g^{\rm s}_{yy}$&$g^{\rm s}_{yz}$&$g^{\rm s}_{zx}$&$g^{\rm s}_{zy}$&$g^{\rm s}_{zz}$  \\
\hline
\hline
0&PM &--&--&--&A&--&--&--&--&--&--&--&--&--&E$_1$&E$_2$&--\\ \hline \hline
\end{tabular}
}
\end{center}
\end{table*}

\subsection{Ordered states}
\label{sec:Ordered states}
Next, we discuss the ASOC additionally induced by spontaneous electronic orders, taking into account the symmetry-breaking term $\tilde{\mathcal{H}}_1$ in equation~(\ref{eq: Ham_MF}) in addition to $\mathcal{H}_0$ in equation~(\ref{Eq:H0}). 
When an electronic order breaks spatial inversion symmetry, the ASOC acquires a spatially uniform component. 
In this section, we discuss the explicit form of such emergent ASOC by taking three typical examples out of the sixteen ordered states: CO (section~\ref{sec:CO}), $zx$-SOO (section~\ref{sec:zx-SOO}), and $xx$-SOO (section~\ref{sec:xx-SOO}). 
The detailed derivation is given in~\ref{Sec:Derivation of Effective Antisymmetric Coupling}. 
We note that staggered components are also induced, but they have the same functional forms as those in the paramagnetic state listed in table~\ref{tab:ASOC_para}; hence, we do not discuss them here. 
The summary including all the other cases will be presented in section~\ref{sec:Summary of Antisymmetric Coupling}. 

\subsubsection{CO (class \#1)}
\label{sec:CO}

In the CO state belonging to the class \#1, $\tilde{\mathcal{H}}_1$ is proportional to $\rho_z \sigma_0 \tau_0$. 
Under this parity breaking order, an additional ASOC is induced in the form of 
\begin{eqnarray}
\label{eq:gCO}
g^{{\rm u}}_{0z}(\bi{k}) \rho_0 \sigma_0 \tau_z \propto -\frac{h t_1^2}{\lambda^2} f_{\rm A}(\bi{k}) \rho_0 \sigma_0 \tau_z. 
\end{eqnarray}
See equation~(\ref{eq:effg}) for the derivation. 
The ASOC is spatially uniform, as it is proportional to $\rho_0$. 
Furthermore, the uniform ASOC is proportional to $h$, which indicates that it is induced by the spontaneous electronic ordering. 
The $\bi{k}$ dependence of the ASOC preserves the threefold rotational symmetry, as discussed in equation~(\ref{eq:f1limit}). 
This is consistent with the fact that the CO state does not break the threefold rotational symmetry, as shown in table~\ref{tab}.

From the viewpoint of symmetry, the effective ASOC in equation~(\ref{eq:gCO}) preserves time-reversal symmetry as $f_{\rm A}(\bi{k})\tau_{z}$ is invariant under the time-reversal operation, while it breaks inversion symmetry. 
This is similar to the Rashba and Dresselhaus SOCs~\cite{rashba1960properties,bychkov1984oscillatory,Dresselhaus_PhysRev.100.580,Dresselhaus_Dresselhaus_Jorio}. 
Indeed, it leads to the spin splitting in the band structure, which is similar to the systems with these SOCs, as discussed in section~\ref{sec:Spin splitting}. 

\subsubsection{$zx$-SOO (class \#5)}
\label{sec:zx-SOO}

Next, we discuss the case of the $zx$-SOO state in the class \#5 ($\tilde{\mathcal{H}}_1 \propto \rho_z \sigma_z \tau_x$).  
In this case, a uniform ASOC is induced in the form of 
\begin{eqnarray}
\label{eq:gzxSOO}
g^{{\rm u}}_{zz}(\bi{k}) \rho_0 \sigma_z \tau_z \propto \frac{h t_0 t_1}{\lambda^2} f_{{\rm E}_1}(\bi{k}) \rho_0 \sigma_z \tau_z. 
\end{eqnarray}
The $\bi{k}$ dependence is different from that in the CO state in equation~(\ref{eq:gCO}): 
the ASOC in the $zx$-SOO state has a linear contribution with respect to $\bi{k}$ as $f_{{\rm E}_1}(\bi{k}) \propto k_y$  in the limit of $\bi{k}\to 0$ [see equation~(\ref{eq:fxylimit})]. 
This is because the threefold rotational symmetry is broken by the $zx$-SOO, as shown in table~\ref{tab}. 
Such a linear term in the ASOC leads to the asymmetric band deformation with respect to $\bi{k}$, as discussed in section~\ref{sec:Band deformation}. 
It also gives rise to the linear magnetoelectric effect, as discussed in section~\ref{sec:Staggered response by electric current}. 

From the symmetry point of view, the effective ASOC in equation~(\ref{eq:gzxSOO}) breaks time-reversal symmetry [$f_{{\rm E}_1}(\bi{k})\sigma_{z}\tau_{z}$ is time-reversal odd] as well as inversion symmetry. 
This indicates that the ASOC induced by $zx$-SOO is regarded as an effective ``toroidal" field along the $k_y$ direction as the $\bi{k}$-linear contribution of $f_{{\rm E}_1}(\bi{k})$, $k_y$, is perpendicular to the spin moment, $\sigma_{z}$. 
In other words, the $zx$-SOO accompanies a ferroic toroidal order in the $k_y$ direction. 
In this situation, there is no guarantee that $k_y$ and $-k_y$ are equivalent, and hence, the asymmetric band deformation is allowed (see section~\ref{sec:Band deformation}).

\subsubsection{$xx$-SOO (class \#7)}
\label{sec:xx-SOO}

The last example discussed here is the $xx$-SOO state in the class \#7 ($\tilde{\mathcal{H}}_1 \propto \rho_z \sigma_x \tau_x$). 
The ordered state in this category breaks all four symmetries, as shown in table~\ref{tab}. 
The emergent ASOC in the $xx$-SOO state is also obtained through equation~(\ref{eq:effg}), whose explicit form is given by 
\begin{eqnarray}
\label{eq:gxxSOO}
g^{{\rm u}}_{xz}(\bi{k}) \rho_0 \sigma_x \tau_z \propto -\frac{h t_0 t_1}{\lambda^2} f_{{\rm E}_1}(\bi{k}) \rho_0 \sigma_x \tau_z. 
\end{eqnarray}
Here, the $\bi{k}$ dependence of the ASOC is linear in $k_y$ in the $\bi{k} \to 0$ limit, similar to that in the $zx$-SOO order, consistent with the breaking of threefold rotational symmetry in the $xx$-SOO. 
Furthermore, the ASOC breaks the mirror symmetry because equation~(\ref{eq:gxxSOO}) includes $\sigma_x$, which does not commute with the mirror operator $\mathcal{M}={\rm i}\sigma_z$ in equation~(\ref{eq:ops}); this is also consistent with the symmetry of the order parameter. 
This form of ASOC induces the uniform longitudinal magnetoelectric effect but no asymmetric band deformation since the $\bi{k}$-linear contribution of $f_{{\rm E}_1}(\bi{k})$, $k_y$, and the spin component, $\sigma_{x}$, are in the $xy$ plane, as discussed in section~\ref{sec:Uniform response by electric current}. 
Similarly, in the $xy$-SOO state, the transverse uniform magnetoelectric effect occurs owing to $g^{\rm u}_{xz} (\bi{k})\propto f_{{\rm E}_2}(\bi{k})$ with an effective ``toroidal'' field along the $k_{z}$ direction.

\subsection{Summary of Antisymmetric Spin-Orbit Coupling}
\label{sec:Summary of Antisymmetric Coupling}

\begin{table*}[htb!]
\begin{center}
\caption{
Emergent uniform ASOCs, $g^{\rm u}_{\alpha \beta}(\bi{k})$ ($\alpha, \beta=0, x, y, z$), in the sixteen ordered states with parity breaking. 
The notations are common to table~\ref{tab:ASOC_para}. 
}
\label{tab:ASOC}
\scalebox{0.85}{
 \begin{tabular}{cccccccccccccccccc}
 \hline
 \hline
\#& 
& $g^{\rm u}_{00}$ & $g^{\rm u}_{x0}$& $g^{\rm u}_{y0}$& $g^{\rm u}_{z0}$  &$g^{\rm u}_{0x}$ &$g^{\rm u}_{0y}$&$g^{\rm u}_{0z}$&$g^{\rm u}_{xx}$&$g^{\rm u}_{xy}$&$g^{\rm u}_{xz}$&$g^{\rm u}_{yx}$&$g^{\rm u}_{yy}$&$g^{\rm u}_{yz}$&$g^{\rm u}_{zx}$&$g^{\rm u}_{zy}$&$g^{\rm u}_{zz}$  \\
\hline
\hline
\multirow{2}{*}{1}&CO&--&--&--&--&--&--&A&--&--&--&--&--&--&--&--&--\\ 
&$zz$-SOO&--&--&--&A&--&--&--&--&--&--&--&--&--&E$_1$&E$_2$&--\\ \hline
\multirow{2}{*}{2}&$x$-OO&--&--&--&--&--&--&E$_1$&--&--&--&--&--&--&--&--&--\\ 
&$y$-OO&--&--&--&--&--&--&E$_2$&--&--&--&--&--&--&--&--&--\\ \hline
\multirow{2}{*}{3}&$xz$-SOO&--&A&--&--&--&--&--&E$_1$&E$_2$&--&--&--&--&--&--&--\\ 
&$yz$-SOO&--&--&A&--&--&--&--&--&--&--&E$_1$&E$_2$&--&--&--&--\\ \hline
\multirow{2}{*}{4}&$z$-SO&--&--&--&--&--&--&--&--&--&--&--&--&--&--&--&A\\ 
&$z$-OO&A&--&--&--&E$_1$&E$_2$&--&--&--&--&--&--&--&--&--&--\\ \hline
\multirow{2}{*}{5}&$zx$-SOO&--&--&--&--&--&--&--&--&--&--&--&--&--&--&--&E$_1$\\ 
&$zy$-SOO&--&--&--&--&--&--&--&--&--&--&--&--&--&--&--&E$_2$\\ \hline
\multirow{2}{*}{6}&$x$-SO&--&--&--&--&--&--&--&--&--&A&--&--&--&--&--&--\\ 
&$y$-SO&--&--&--&--&--&--&--&--&--&--&--&--&A&--&--&--\\ \hline
\multirow{4}{*}{7}&$xx$-SOO&--&--&--&--&--&--&--&--&--&E$_1$&--&--&--&--&--&--\\ 
&$yy$-SOO&--&--&--&--&--&--&--&--&--&--&--&--&E$_2$&--&--&--\\ 
&$xy$-SOO&--&--&--&--&--&--&--&--&--&E$_2$&--&--&--&--&--&--\\ 
&$yx$-SOO&--&--&--&--&--&--&--&--&--&--&--&--&E$_1$&--&--&--\\ \hline \hline
\end{tabular}
}
\end{center}
\end{table*}

Performing similar analysis for other ordered states, we obtain the uniform ASOCs induced by each order. 
Table~\ref{tab:ASOC} summarizes the results, in which A, E$_1$, and E$_2$ represent that the induced uniform ASOC has the $\bi{k}$ dependence of $f_{\rm A}(\bi{k})$, $f_{\rm E_{1}}(\bi{k})$, and $f_{\rm E_{2}}(\bi{k})$, respectively. 
For example, in the case of CO in the class \#1 described in section~\ref{sec:CO}, the system exhibits the uniform ASOC given by $g^{\rm u}_{0z}(\bi{k})\rho_0 \sigma_0 \tau_z \propto f_{\rm A}(\bi{k})\rho_0 \sigma_0 \tau_z$, while the $zx$-SOO in the class \#5 in section~\ref{sec:zx-SOO} induces $g^{\rm u}_{zz}(\bi{k})\rho_0 \sigma_z \tau_z \propto f_{\rm E_{1}}(\bi{k})\rho_0 \sigma_z \tau_z$. 

Table~\ref{tab:ASOC} shows that a variety of uniform ASOCs are induced by spontaneous parity breaking. 
They have different spin and orbital dependences according to the types of ordered phases, which are useful for understanding of peculiar electronic structures under each electronic order. 
Moreover, table~\ref{tab:ASOC} is also useful for understanding of the physical properties in each phase, such as the magnetoelectric effects, discussed in section~\ref{sec:Cross Correlation under Antisymmetric Coupling}.

\section{Electronic Structure}
\label{sec:Electronic Structure}
In this section, we show the electronic band structures of the paramagnetic and ordered states classified into eight classes \#0-7 in tables~\ref{tab}, \ref{tab:ASOC_para}, and \ref{tab:ASOC}. 
We discuss the relationship between the ASOCs induced by electronic ordering and the band structures. 
After briefly introducing the typical band structure in the paramagnetic state (class \#0) in section~\ref{sec:Band structure in paramagnetic state}, we show the spin splitting in the classes \#1, \#2, and \#3 (section~\ref{sec:Spin splitting}), valley splitting in the class \#4 (section~\ref{sec:Valley splitting}), and asymmetric band deformation with a band bottom shift in the class \#5 (section~\ref{sec:Band deformation}) in the electronic structures. 
We also present the complementary understanding of the peculiar band modulations from the eigenvalues of the Hamiltonian in \ref{sec:Comparison between the antisymmetric spin-orbit coupling and eigenvalues}. 
Hereafter, we take $t_0=0.5$, $t_1=0.5$, and $\lambda=0.5$ if not explicitly stated.  

\subsection{Band structure in the paramagnetic state (class \#0)}
\label{sec:Band structure in paramagnetic state}

\begin{figure}[htb!]
\centering
\includegraphics[width=0.75 \hsize]{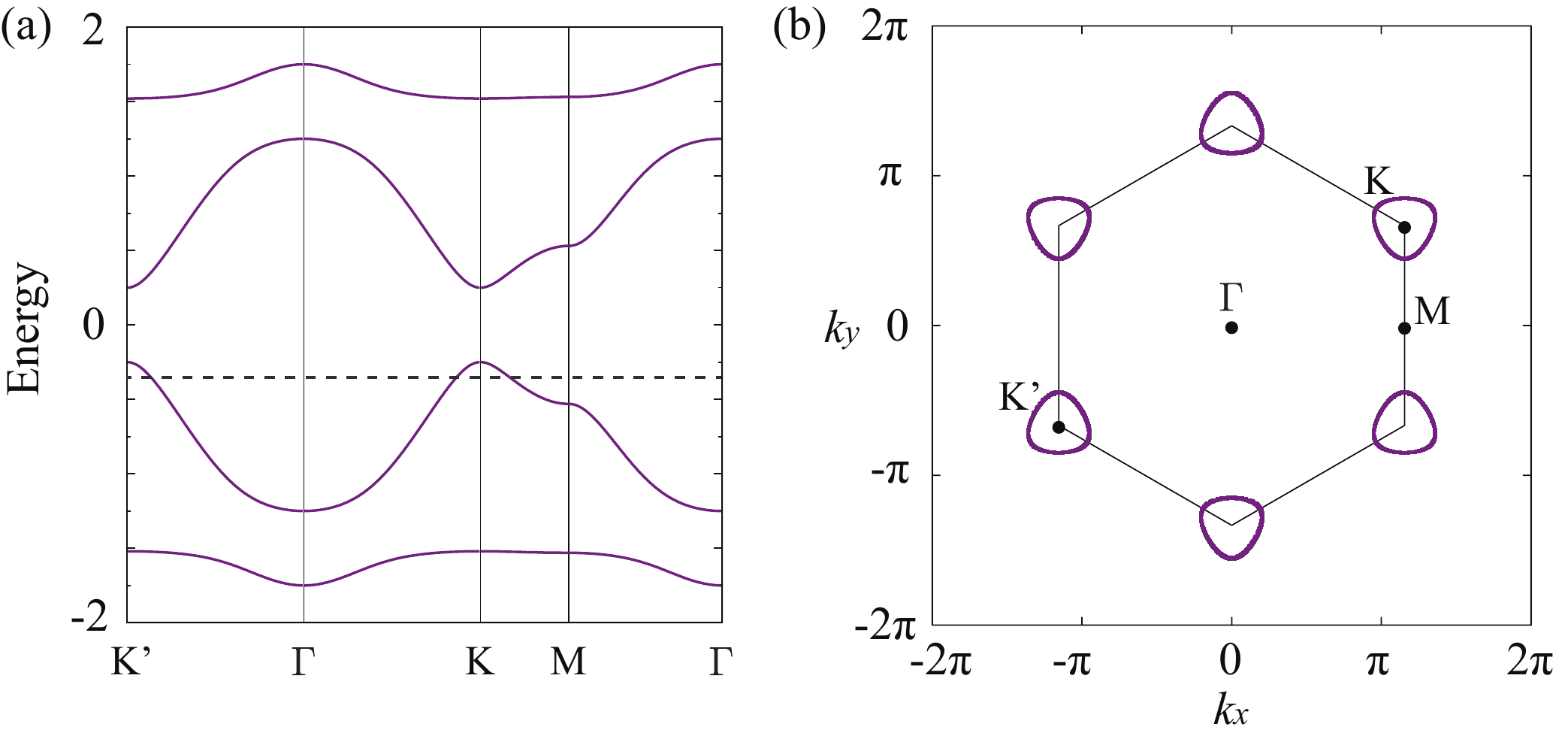} 
\caption{
\label{fig:band_para}
(a) Electronic band structure for the paramagnetic state. 
(b) The energy contours slightly below the Fermi level at half filling ($E=-0.35$) 
corresponding to the dashed line in (a). 
The hexagon represents the Brillouin zone. 
}
\end{figure}

Figure~\ref{fig:band_para} shows the band structure in the paramagnetic state, calculated from the noninteracting Hamiltonian $\mathcal{H}_0$ in equation~(\ref{Eq:H0}). 
In the paramagnetic state, there are four bands separated by energy gaps due to both the atomic SOC $\lambda$ and inter-orbital hopping $t_1$. 
Each band is doubly degenerate as both spatial inversion and time-reversal symmetries are preserved. 
At the commensurate fillings, $n_{\rm e}=1$, $2$, and $3$, the system becomes insulating. 
These insulating states are topological insulators; the spin Hall conductivity [see equation~(\ref{eq:sigmaxy_spin})] is quantized at a nonzero integer value, as mentioned above~\cite{Hayami_1742-6596-592-1-012131}.  
This is due to the presence of the staggered ASOCs, $g^{{\rm s}}_{z0}(\bi{k})$, $g^{{\rm s}}_{zx}(\bi{k})$ and $g^{{\rm s}}_{zy}(\bi{k})$, in table~\ref{tab:ASOC_para}. 

The energy contours slightly below the Fermi level at half filling is also shown in figure~\ref{fig:band_para}(b). 
Reflecting the presence of spatial inversion and threefold rotational symmetries, the six regions near the K and K' points in the Brillouin zone are all equivalent. 
Such degeneracy is lifted once the electronic ordering breaks the symmetries, as discussed in the following sections. 

\subsection{Spin splitting (class \#1, \#2, \#3)}
\label{sec:Spin splitting}
In this section, we show that particular parity breaking orders split the electronic bands depending on the spin degree of freedom, called spin splitting. 
The spin splitting occurs in the classes \#1, \#2, and \#3 (SS in table~\ref{tab}), but in a different form in each class. 
Note that the spin splitting occurs only in the presence of time-reversal symmetry. 

\begin{figure}[htb!]
\centering
\includegraphics[width=0.8\hsize]{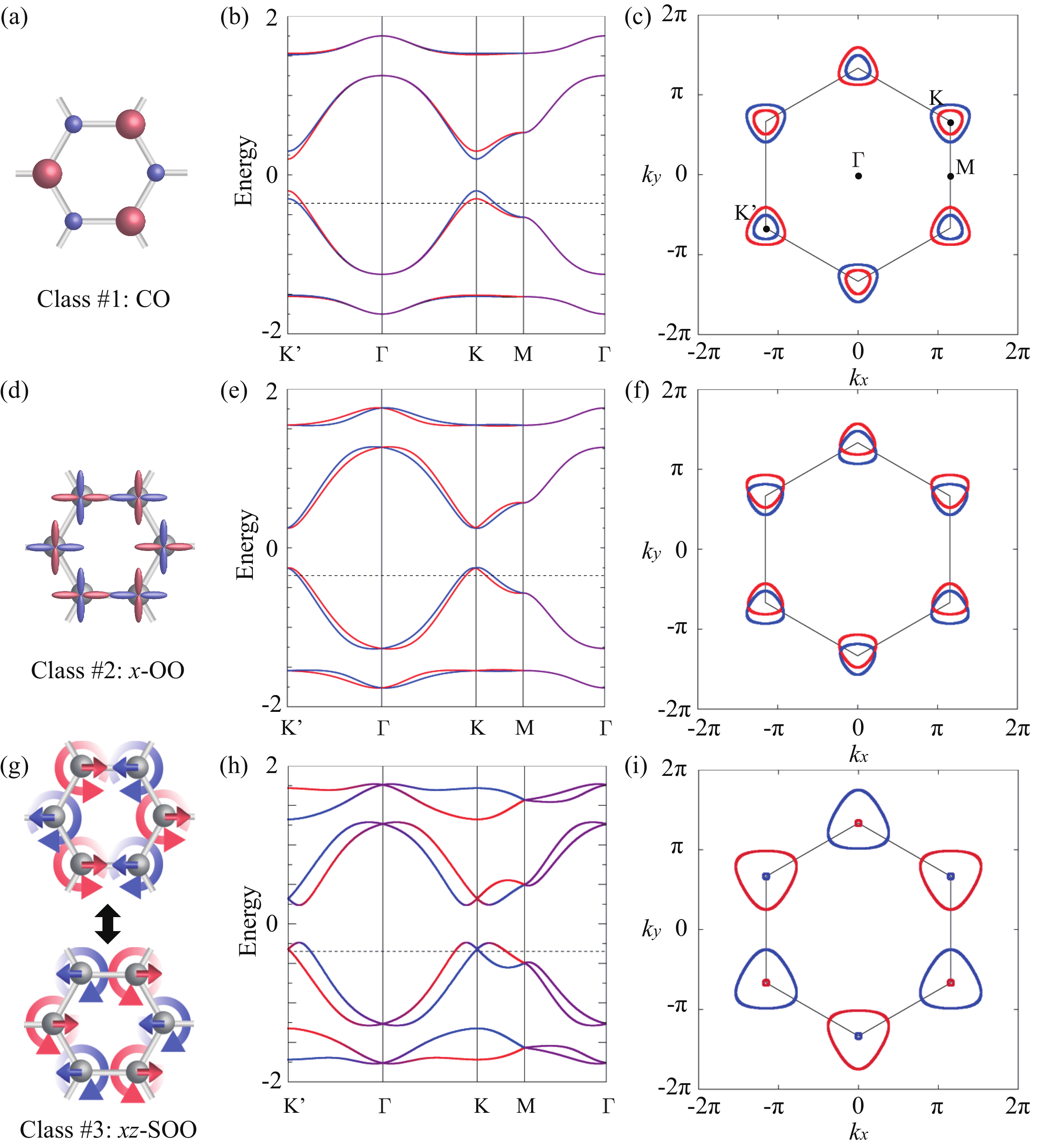} 
\caption{
\label{fig:band_splitting}
The schematic pictures of the charge, spin, and orbital order in (a) CO, (d) $x$-OO, and (g) $xz$-SOO phases are shown. 
In the picture for CO in (a), the size of the spheres reflects the magnitude of local charge densities. 
For $x$-OO in (d), the lobes represent the $d_{x^2-y^2}$ orbitals. 
For $xz$-SOO in (g), the arrows on the spheres show the local magnetic moments along the $x$ direction and the circular arrows around the spheres represent the orbital currents. 
The $xz$-SOO state is given by the superposition of the two configurations. 
Electronic band structures of the mean-field Hamiltonian ${\cal H}_0 + \tilde{{\cal H}}_1$ for the (b) CO, (e) $x$-OO, and (h) $xz$-SOO states. 
The energy contours slightly below the Fermi level at half filling ($E=-0.35$) corresponding to the dashed lines in (b), (e), and (h) are shown in (c), (f), and (i), respectively. 
We take $h=0.05$ ($0.2$) for (b) and (c) [(e), (f), (h), and (i)]. 
In (b), (c), (e), and (f), the red (blue) curves show the bands with up(down)-spin polarization in the $z$ direction, while those are for up(down)-spin polarization in the $x$ direction in (h) and (i). 
}
\end{figure}

\subsubsection{Class \#1}
\label{sec:Class1}
First, we discuss the electronic structure in the CO state classified into the class \#1. 
Figure~\ref{fig:band_splitting}(a) shows a schematic picture of CO. 
The order parameter in the CO state breaks only spatial inversion symmetry, as shown in table~\ref{tab}. 
Figure~\ref{fig:band_splitting}(b) shows the band structure in the CO state calculated from the Hamiltonian $\mathcal{H}_0 + \tilde{\mathcal{H}}_1$ at $h=0.05$. 
The red (blue) curves denote the up(down)-spin component, which indicates that the spin splitting occurs in the CO state. 
The spin splitting is antisymmetric with respect to the wave vector $\bi{k}$: the low energy states in both valence and conduction bands at half filling are predominantly spin-up polarized at the K' point, while they are spin-down polarized at the K point.
This is also seen in the energy contours slightly below the Fermi level at half filling, as shown in figure~\ref{fig:band_splitting}(c). 
The antisymmetric spin splitting is large around the K and K' points, as shown in figure~\ref{fig:band_splitting}(b). 
This is consistent with the fact that the ASOC in equation~(\ref{eq:gCO}) does not include the linear term with respect to the wave vector, but the third-order term [see also equation~(\ref{eq:f1limit})]. 

It is also interesting to examine the electronic structure from the topological aspect. 
For this purpose, we compute the spin Hall conductivity by using the linear response theory:  
\begin{eqnarray}
\label{eq:sigmaxy_spin}
\sigma_{xy}^{{\rm SH}} =-\frac{e}{2} \frac{1}{{\rm i} V_0} \sum_{m n \bi{k}} \frac{f(\varepsilon_{n \bi{k}})-f(\varepsilon_{m \bi{k}})}{\varepsilon_{n \bi{k}}-\varepsilon_{m \bi{k}}} \frac{J^{(s) n m
}_{x,\bi{k}} J^{ m n}_{y,\bi{k}}}{\varepsilon_{n \bi{k}}-\varepsilon_{m \bi{k}} + {\rm i} \delta}, 
\end{eqnarray} 
where $V_{0}$ is the system volume, $f(\varepsilon)$ is the Fermi distribution function, and
$\varepsilon_{m\bi{k}}$ and $|m\bi{k} \rangle$ are the eigenvalue and eigenstate of ${\cal H}_0+\tilde{{\cal H}}_1$. 
$J_{\nu, \bi{k}}^{(s) m n} = \langle m \bi{k} | J_{\nu}^{({\rm s})}| n \bi{k} \rangle$ is the matrix element of the spin current operator, which is defined by $J_\nu^{({ s})}= \frac{1}{2}\{\sigma_z, J_{\nu}\}$ in the $\nu$ direction ($J_{\nu}$ is the current operator and $\{ \cdots \}$ is an anticommutator). 
We set $-e/4\pi=1$ ($e$ is the elementary charge). 
Thus, $\sigma_{xy}^{{\rm SH}}$ represents the coefficient for the spin current in the $x$ direction induced by the electric current in the $y$ direction~\footnote{In the case of insulators, the electric current should be replaced with the electric field. This is also the case for the off-diagonal responses in equation~(\ref{Eq:Mangetoelectric effect_mae}).}. 
We take temperature $T=0.001$ and the damping factor $\delta=0.001$. 

\begin{figure}[htb!]
\centering
\includegraphics[width=0.75 \hsize]{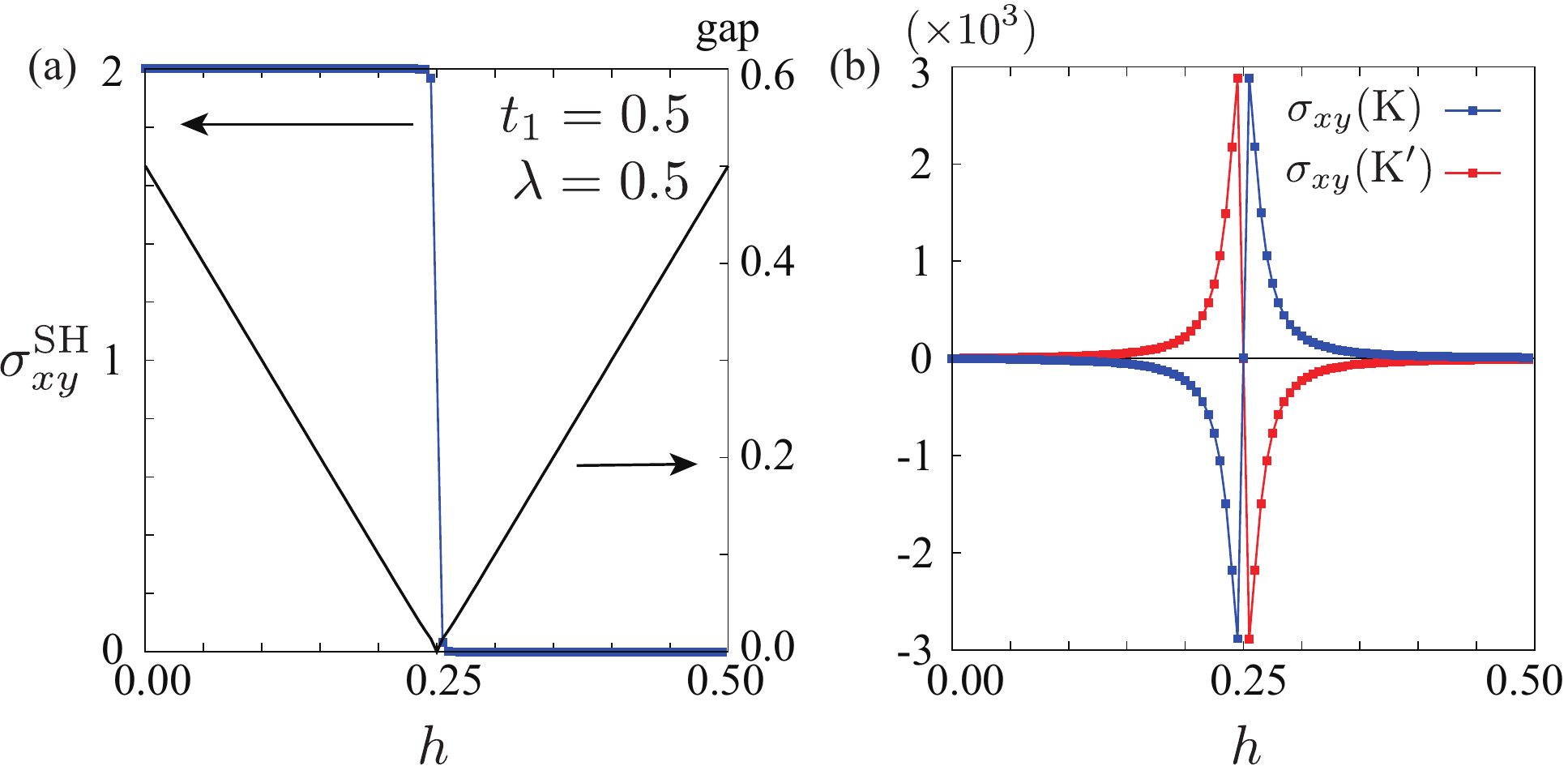} 
\caption{
\label{fig:spinsplitting_Hall}
(a) $h$ dependence of the spin Hall conductivity at half filling in the CO state (left axis). 
The slight deviations from integer values are due to the finite-size effects. 
The energy gap is also shown (the right axis). 
(b) $h$ dependences of the Hall conductivities at the K and K' points. 
}
\end{figure}

Figure~\ref{fig:spinsplitting_Hall}(a) shows the $h$ dependence of the spin Hall conductivity in the CO state. 
For $h < \lambda/2=0.25$, the spin Hall conductivity $\sigma_{xy}^{{\rm SH}}$ is quantized at 2, indicating that the system is a topological insulator~\cite{Hayami_1742-6596-592-1-012131}. 
With increasing $h$, the band gap shrinks as shown in figures~\ref{fig:spinsplitting_Hall}(a) [see also figure~\ref{fig:energylevel_CO}(a) in \ref{sec:Comparison between the antisymmetric spin-orbit coupling and eigenvalues}], and it closes at the K and K' points at $h = \lambda/2$, where $\sigma_{xy}^{{\rm SH}}$ changes discontinuously from 2 to 0. 
For further increasing $h$, the gap opens again, but $\sigma_{xy}^{{\rm SH}}$ remains at 0, as shown in figure~\ref{fig:spinsplitting_Hall}(a): the system is a trivial band insulator for $h>\lambda/2$. 
The result indicates the transition from the topological insulator to a trivial band insulator at $h=\lambda/2$. 

On the other hand, the CO state also exhibits the valley Hall effect, reflecting the emergence of the valley degree of freedom (inequivalence between the K and K' points)~\cite{Xiao_PhysRevLett.99.236809}. 
Figure~\ref{fig:spinsplitting_Hall}(b) shows $h$ dependences of the Hall coefficients $\sigma_{xy}$ at the K and K' points. 
Here, $\sigma_{xy}$ is calculated from the correlation between the electric currents $J_x$ and $J_y$ within the linear response theory similar to equation~(\ref{eq:sigmaxy_spin}): 
\begin{eqnarray}
\label{eq:sigmaxy_valley}
\sigma_{xy} (\bkappa) =\frac{e^2}{ \hbar} \frac{1}{{\rm i}} \sum_{m n} \frac{f(\varepsilon_{n \bkappa})-f(\varepsilon_{m \bkappa})}{\varepsilon_{n \bkappa}-\varepsilon_{m \bkappa}} \frac{J^{ n m
}_{x,\bkappa} J^{ m n}_{y,\bkappa}}{\varepsilon_{n \bkappa}-\varepsilon_{m \bkappa} + {\rm i} \delta}, 
\end{eqnarray} 
where $\hbar$ is the Dirac constant and $\bkappa$ stands for either the K or K' points.
As shown in figure~\ref{fig:spinsplitting_Hall}(b), the Hall conductivities at the K and K' points become nonzero with opposite signs due to the presence of time-reversal symmetry. 
They are critically enhanced with the inverse square of the energy gap while approaching the gap closing point at $h=\lambda/2$ from both sides. 
The result indicates that we can obtain gigantic valley Hall responses by controlling the order parameter in the CO phase. 

The antisymmetric spin splitting also occurs for the $zz$-SOO phase in the same class \#1, but in a different form: the energy levels just above and below the Fermi level at half filling at the K point have opposite spin polarizations. 
The difference is explained by the spin dependence of the emergent ASOCs for the CO and $zz$-SOO states; the former is proportional to $\sigma_{0}\tau_{z}$ which affects the spin sector via the spin-orbit coupling $\propto\sigma_{z}\tau_{z}$, whereas the latter is directly proportional to $\sigma_z\tau_{0}$ ($g^{{\rm u}}_{z0}$), as shown in table~\ref{tab:ASOC}. 

\subsubsection{Class \#2}
\label{sec:Class2}
Next, we turn to the class \#2. 
The $x/y$-OO states in this class are characterized by the simultaneous breaking of spatial inversion and rotational symmetries, as shown in table~\ref{tab}. 
In this situation, a uniform ASOC with $\bi{k}$-linear contribution appears, as shown in table~\ref{tab:ASOC}. 
In the $x$-OO case, whose schematic picture is shown in figure~\ref{fig:band_splitting}(d), we obtain the ASOC from equation~(\ref{eq:effg}) in the form of 
 \begin{eqnarray}
\label{eq:g0xOO}
g^{{\rm u}}_{0z}(\bi{k})\rho_0 \sigma_0 \tau_z \propto \frac{h t_0 t_1}{\lambda^2} f_{{\rm E}_1}(\bi{k}) \rho_0 \sigma_0 \tau_z, 
\end{eqnarray}
which is proportional to $k_y$ in the limit of $\bi{k} \to 0$, as shown in equation~(\ref{eq:fxylimit}). 
On the other hand, the ASOC for $y$-OO is given as
\begin{eqnarray}
\label{eq:g0yOO}
g^{{\rm u}}_{0z}(\bi{k}) \rho_0 \sigma_0 \tau_z \propto \frac{h t_0 t_1}{\lambda^2} f_{{\rm E}_2}(\bi{k}) \rho_0 \sigma_0 \tau_z,   
\end{eqnarray}
which is proportional to $k_x$ in the limit of $\bi{k} \to 0$. 

The $\bi{k}$-linear dependence of the emergent ASOC in the class \#2 leads to a different type of spin splitting from the class \#1.
For instance, in the $x$-OO phase, the $k_y$-linear contribution brings about an antisymmetric spin splitting occurs in a way that the bands with up(down)-spin polarization are shifted to the $+k_y$ ($-k_y$) direction. 
The typical band structure is shown in figure~\ref{fig:band_splitting}(e). 
Note that the band dispersion still satisfies the relation $\varepsilon_{\sigma} (\bi{k})=\varepsilon_{-\sigma} (-\bi{k})$ due to the presence of time-reversal symmetry. 
As seen in the energy contours plotted in figure~\ref{fig:band_splitting}(f), however, the band structure is no longer symmetric with respect to the threefold rotation. 
Another distinct feature from the class \#1 is the $\bi{k}$ dependence of the magnitude of the antisymmetric spin splitting. 
The spin splitting for the present $x$-OO order takes place predominantly around the $\Gamma$ point rather than the K and K' points as shown in figure~\ref{fig:band_splitting}(e), 
whereas that for the CO appears conspicuously around the K and K' points as shown in figure~\ref{fig:band_splitting}(b). 
This is because the lowest-order ASOC is linear in $\bi{k}$ in the class \#2. 
The situation is similar to the other state in this class \#2, the $y$-OO phase, while the bands are split in the $k_x$ direction reflecting the $k_x$-linear contribution in the ASOC in equation~(\ref{eq:g0yOO}). 

\subsubsection{Class \#3}
\label{sec:Class3}
Finally, let us discuss the class \#3. 
The order parameter in this class breaks spatial inversion and mirror symmetries but it retains rotational symmetry, as shown in table~\ref{tab}. 
In this case also, the band structure exhibits spin splitting: figures~\ref{fig:band_splitting}(h) and \ref{fig:band_splitting}(i) show the typical examples under $xz$-SOO whose schematic picture is shown in figure~\ref{fig:band_splitting}(g). 
However, the spin polarization is not along the $z$ axis but in the $xy$ plane. 
The spin splitting is understood from the form of the induced ASOC. 
For instance, in the $xz$-SOO state, according to table~\ref{tab:ASOC}, the induced ASOCs are summarized into the form, 
\begin{equation}
\left[c_{1}f_{\rm A}(\bi{k})\tau_{0}+c_{2}\{f_{{\rm E}_1}(\bi{k})\tau_{x}+f_{{\rm E}_2}(\bi{k})\tau_{y}\}\right]\sigma_{x}\rho_{0}.
\label{eq:g0class3}
\end{equation}
The form of emergent ASOC is similar to that in the $zz$-SOO in the class \#1, while the spin component is $\sigma_x$ instead of $\sigma_z$. 
Thus, the ASOC induced by the $xz$-SOO state leads to the spin splitting with the spin quantization axis along the $x$ direction, as shown in figures~\ref{fig:band_splitting}(h) and \ref{fig:band_splitting}(i). 
Similarly, the spin splitting along the $\sigma_y$ direction is obtained for the $yz$-SOO phase in the same class. 

\subsection{Valley splitting (class \#4)}
\label{sec:Valley splitting}

\begin{figure}[htb!]
\centering
\includegraphics[width=0.75 \hsize]{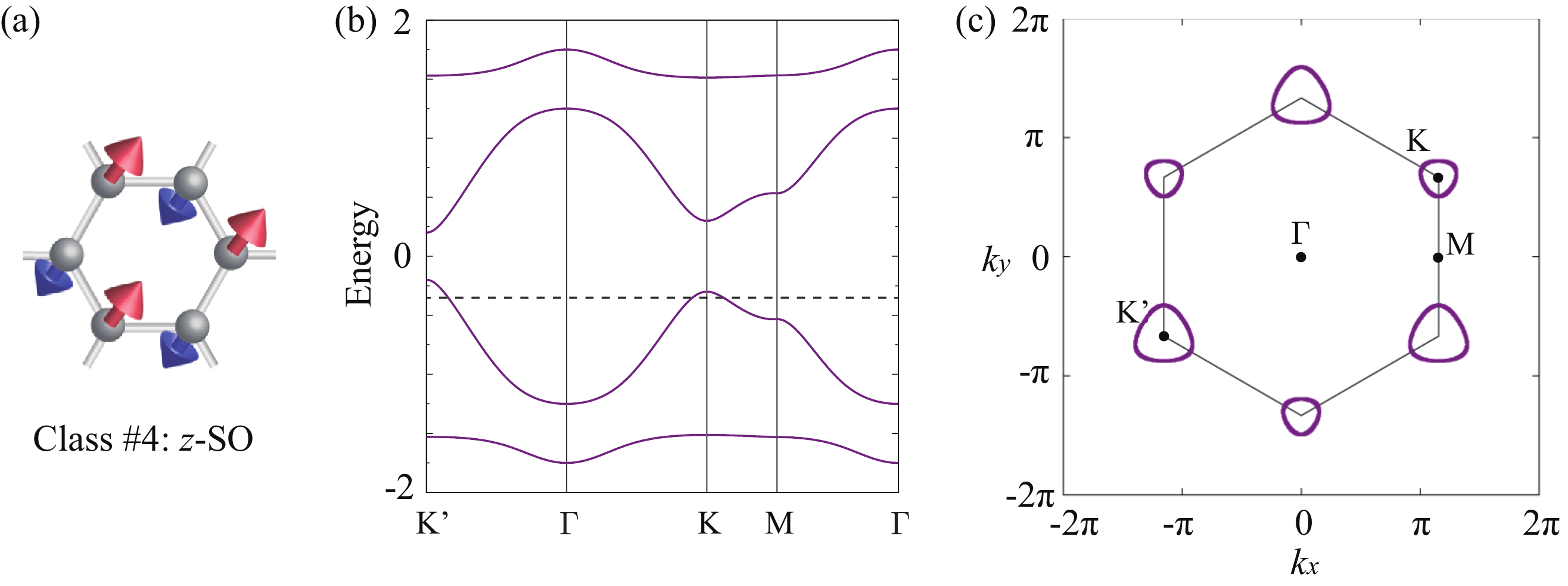} 
\caption{
\label{fig:band_valleysplitting}
(a) The schematic picture for $z$-SO is shown. 
(b) Electronic band structure of the mean-field Hamiltonian ${\cal H}_0 + \tilde{{\cal H}}_1$ for the $z$-SO state at $h=0.05$. 
(c) The energy contours slightly below the Fermi level at half filling ($E=-0.35$) 
corresponding to the dashed line in (b). 

}
\end{figure}

In this section, we show that the parity breaking order in the class \#4 leads to a different type of modulation of the band structure. 
In this case, the spin splitting is absent, but the bands are modulated in a different way between the K and K' points. 
This is called valley splitting (VS in table~\ref{tab}). 

The $z$-SO and $z$-OO belonging to the class \#4 both break spatial inversion and time-reversal symmetries simultaneously, as shown in table~\ref{tab}. 
Hence, there is no guarantee that the energy eigenvalue at $\bi{k}$ is degenerate with that at $-\bi{k}$, as in the toroidal ordered state~\cite{Yanase:JPSJ.83.014703,Hayami_PhysRevB.90.024432}. 
In fact, these orders can be regarded as toroidal octupole orders (see section~\ref{sec:odd-parity}), and the lowest-order contribution of the emergent ASOC is of third order with respect to $\bi{k}$ (see table~\ref{tab:ASOC})~\footnote{The ASOCs with $\bi{k}$-linear component do not break the rotational symmetry of the system because the net component given by their linear combination preserves the rotational symmetry, similar to the discussion in Sec.~\ref{sec:Paramagnetic state}.}. 

Specifically, in the $z$-SO phase in figure~\ref{fig:band_valleysplitting}(a), the band structure is modulated as shown in figure~\ref{fig:band_valleysplitting}(b): at half filling, the gap at the K' point becomes smaller than that at the K point. 
Correspondingly, the hole pockets at the K' point are larger than those at the K points, as shown in the energy contours in figure~\ref{fig:band_valleysplitting}(c). 
Thus, the $z$-SO leads to valley splitting in the band structure, as pointed out in~\cite{li2013coupling}.  
This valley splitting is induced by the ASOC generated by the $z$-SO, which is represented by 
\begin{eqnarray}
\label{eq:gzSO}
g^{\rm u}_{zz}(\bi{k})\rho_0 \sigma_z \tau_z \propto -\frac{h t_1^2}{\lambda^2} f_{\rm A}(\bi{k}) \rho_0 \sigma_z \tau_z. 
\end{eqnarray}
The form of the ASOC is similar to that in the CO state in equation~(\ref{eq:gCO}); the difference is in the spin component, $\sigma_0 \to \sigma_z$, reflecting the time-reversal symmetry breaking. 
The valley splitting is understood by the appearance of $f_{\rm A}(\bi{k})$ due to the rotational symmetry and the fact that $f_{\rm A}(\bi{k})\sigma_{z}\tau_{z}$ is invariant under simultaneous transformations of $\mathcal{PT}$ (see also \ref{sec:Comparison between the antisymmetric spin-orbit coupling and eigenvalues}). 
Furthermore, the spin Hall conductivity also shows the similar behavior to that in the class \#1, as shown in figure~\ref{fig:spinsplitting_Hall}(a): the similar topological transition takes place at $h=\lambda/2$. 

The valley splitting is also seen in the $z$-OO phase in the same class, but in a different way from the $z$-SO case. 
In the $z$-OO phase, for example, the energy levels just above and below the Fermi level at half filling are shifted downward (upward) at the K' (K) point. 
This is due to the spin and orbital dependence of the emergent ASOC, $g^{\rm u}_{00}(\bi{k})\rho_0\sigma_0\tau_0$, as listed in table~\ref{tab:ASOC}.  

\subsection{Asymmetric band deformation (class \#5)}
\label{sec:Band deformation}

\begin{figure}[htb!]
\centering
\includegraphics[width=0.75 \hsize]{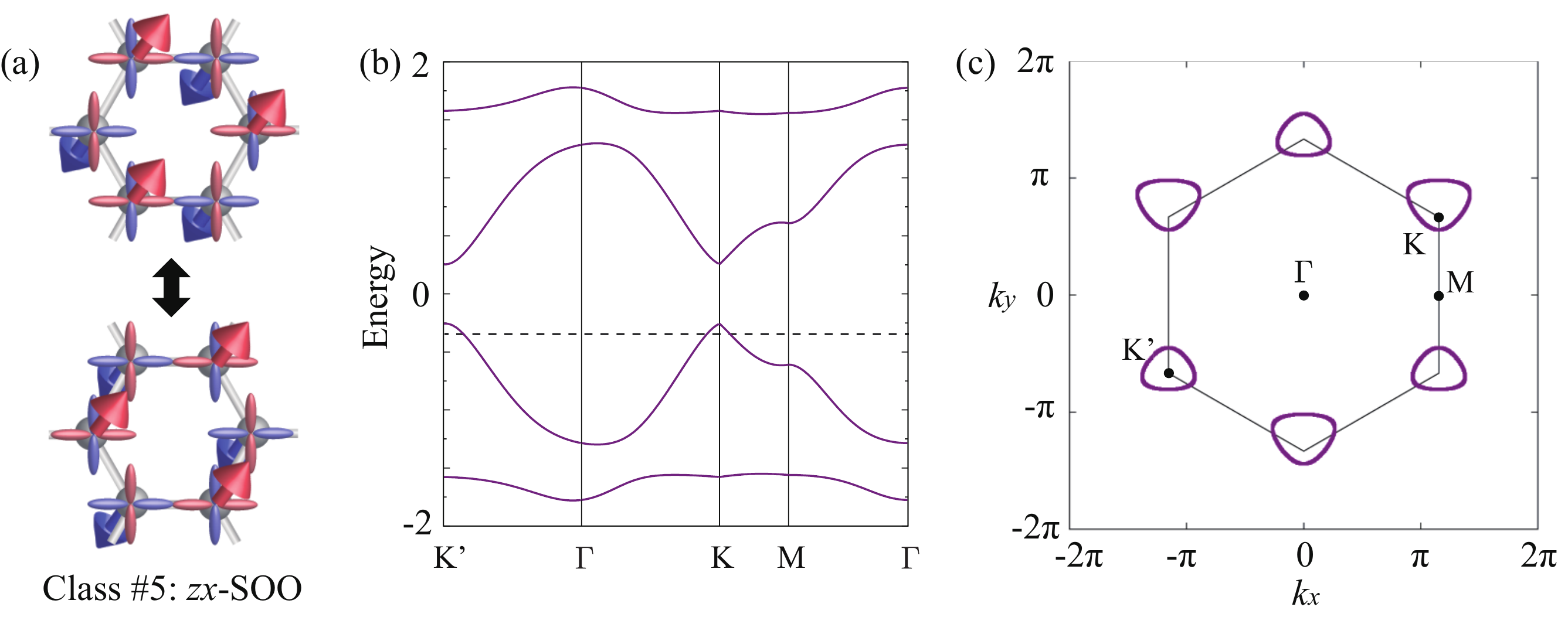} 
\caption{
\label{fig:band_deformation}
(a) The schematic picture for $zx$-SOO is shown. 
(b) Electronic band structure of the mean-field Hamiltonian ${\cal H}_0 + \tilde{{\cal H}}_1$ for the $zx$-SOO state at $h=0.3$. 
(c) The energy contours slightly below the Fermi level at half filling ($E=-0.35$) corresponding to the dashed line in (b). 
}
\end{figure}

Finally, we discuss the electronic band structure in the presence of parity breaking orders classified in the class \#5, which break spatial inversion, time-reversal, and rotational symmetries, as shown in table~\ref{tab}. 
In this case, for instance, in the $zx$-SOO state schematically shown in figure~\ref{fig:band_deformation}(a), the emergent ASOC includes the $k_y$-linear contribution as inferred by equation~(\ref{eq:gzxSOO}), similar to the case of the $x$-OO state in the class \#2 [equation~(\ref{eq:g0xOO})]. 
Thus, we expect the asymmetric band deformation in the $k_y$ direction. 
The difference between the two classes is the spin dependence: the former includes $\sigma_z$, while the latter $\sigma_0$. 
This leads to the different type of the asymmetric band deformation in the class \#5, as demonstrated in figures~\ref{fig:band_deformation}(b) and \ref{fig:band_deformation}(c): the bands are modulated with a band bottom shift, with retaining the spin degeneracy as $f_{{\rm E}_1}(\bi{k})\sigma_{z}\tau_{z}$ is invariant under $\mathcal{PT}$. 
The band deformation is similar to the ferroic toroidal ordered cases discussed in \cite{Yanase:JPSJ.83.014703,Hayami_PhysRevB.90.024432,Hayami_doi:10.7566/JPSJ.84.064717}.  
In the case of the $zy$-SOO phase in the same class \#5, a similar band deformation takes place with the band bottom shift to the $k_x$ direction because of the induced ASOC, $g^{{\rm u}}_{zz}(\bi{k})  \propto k_x$. 

\section{Off-diagonal Responses}
\label{sec:Cross Correlation under Antisymmetric Coupling}
In this section, we discuss the off-diagonal responses of the paramagnetic and ordered states classified into eight classes. 
Specifically, we show the spatially uniform and staggered moments induced by an electric current in sections~\ref{sec:Uniform response by electric current} and \ref{sec:Staggered response by electric current}, respectively.  
We use the linear response theory by computing the tensors between order parameters and electric current in each ordered state:
\begin{eqnarray}
K^{\rm T}_{\alpha\beta\mu} 
=\frac{
2\pi}{{\rm i} V_{0}} \sum_{m n \bi{k}} \frac{f(\varepsilon_{
n\bi{k}})-f(\varepsilon_{m\bi{k}})}{\varepsilon_{n\bi{k}}-\varepsilon_{ m\bi{k}}} 
\frac{m_{{\rm T}\alpha \beta,\bi{k}}^{nm} J_{\mu,\bi{k}}^{mn} 
}{\varepsilon_{n\bi{k}}-\varepsilon_{m\bi{k}}+ {\rm i} \delta}, 
\label{Eq:Mangetoelectric effect_mae}
\end{eqnarray}
where $m_{{\rm T}\alpha\beta,\bi{k}}^{n m}=\langle n\bi{k} | \rho_{\rm T}\Lambda^{\alpha}_{\;\beta}  | m\bi{k} \rangle$ 
[${\rm T}={\rm u}$ ($\rho_{\rm u}=\rho_0$) or s ($\rho_{\rm s}=\rho_z$)]. 
Thus, $K^{\rm u}_{\alpha\beta\mu}$ ($K^{\rm s}_{\alpha\beta\mu}$) represents the coefficient for the uniform (staggered) order parameter $\propto \sigma_{\alpha} \tau_{\beta}$ induced by the electric current in the $\mu$ direction. 
In equation~(\ref{Eq:Mangetoelectric effect_mae}), we set $g \mu_{{\rm B}}e/(2h)=1$ ($g$ is the $g$-factor and $\mu_{{\rm B}}$ the Bohr magneton). 

\subsection{Uniform response to electric current}
\label{sec:Uniform response by electric current}

\begin{table*}[htb!]
\begin{center}
\caption{
Uniform off-diagonal responses in paramagnetic and sixteen staggered ordered states. 
C, S$_{\alpha}$, O$_{\beta}$, and SO$_{\alpha\beta}$ ($\alpha,\beta= x, y, z$) represent the charge, spin, orbital, and spin-orbital orders, corresponding to the coefficients $K^{{\rm u}}_{00\mu}$, $K^{{\rm u}}_{\alpha0\mu}$, $K^{{\rm u}}_{0\beta\mu}$, and $K^{{\rm u}}_{\alpha\beta\mu}$, respectively. 
$\mu=x,y$ in the table represents that the coefficient becomes nonzero for the electric current along the $\mu$ direction. 
The superscript $^*$ indicates the nonzero response even in the absence of SOC, $\lambda=0$. 
}
\label{tab:MEuniform}
\scalebox{0.76}{
 \begin{tabular}{cccccccccccccccccc}
 \hline
\#& & C & ${\rm S}_x$& ${\rm S}_y$& ${\rm S}_z$  &${\rm O}_x$ &${\rm O}_y$&${\rm O}_z$&${\rm SO}_{xx}$&${\rm SO}_{xy}$&${\rm SO}_{xz}$&${\rm SO}_{yx}$&${\rm SO}_{yy}$&${\rm SO}_{yz}$&${\rm SO}_{zx}$&${\rm SO}_{zy}$&${\rm SO}_{zz}$  \\
\hline
\hline
0&PM           &--&--&--&--&--&--&--&--&--&--&--&--&--&--&--&--\\ \hline
\multirow{2}{*}{1}&CO           &--&--&--&--&$x^*$&$y^*$&--&--&--&--&--&--&--&$y$&$x$&--\\ 
&$zz$-SOO&--&--&--&--&$x$&$y$&--&--&--&--&--&--&--&$y^*$&$x^*$&--\\ \hline
\multirow{2}{*}{2}&$x$-OO    &--&--&--&--&$x^*$&$y^*$&$y^*$&--&--&--&--&--&--&$y$&$x$&$x$\\ 
&$y$-OO    &--&--&--&--&$y^*$&$x^*$&$x^*$&--&--&--&--&--&--&$x$&$y$&$y$\\ \hline
\multirow{2}{*}{3}&$xz$-SOO&--&--&--&--&--&--&--&$y^*$&$x^*$&--&$x$&$y$&--&--&--&--\\ 
&$yz$-SOO&--&--&--&--&--&--&--&$x$&$y$&--&$y^*$&$x^*$&--&--&--&--\\ \hline
\multirow{2}{*}{4}&$z$-SO    &--&--&--&--&$y$&$x$&--&--&--&--&--&--&--&$x^*$&$y^*$&--\\ 
&$z$-OO    &--&--&--&--&$y^*$&$x^*$&--&--&--&--&--&--&--&$x$&$y$&--\\ \hline
\multirow{2}{*}{5}&$zx$-SOO&--&--&--&--&$y$&$x$&$x$&--&--&--&--&--&--&$x^*$&$y^*$&$y^*$\\ 
&$zy$-SOO&--&--&--&--&$x$&$y$&$y$&--&--&--&--&--&--&$y^*$&$x^*$&$x^*$\\ \hline
\multirow{2}{*}{6}&$x$-SO    &--&--&--&--&--&--&--&$x^*$&$y^*$&--&$y$&$x$&--&--&--&--\\ 
&$y$-SO    &--&--&--&--&--&--&--&$y$&$x$&--&$x^*$&$y^*$&--&--&--&--\\ \hline
\multirow{4}{*}{7}&$xx$-SOO&--&$x$&$y$&--&--&--&--&$x^*$&$y^*$&$y^*$&$y$&$x$&$x$&--&--&--\\ 
&$yy$-SOO&--&$x$&$y$&--&--&--&--&$x$&$y$&$y$&$y^*$&$x^*$&$x^*$&--&--&--\\ 
&$xy$-SOO&--&$y$&$x$&--&--&--&--&$y^*$&$x^*$&$x^*$&$x$&$y$&$y$&--&--&--\\ 
&$yx$-SOO&--&$y$&$x$&--&--&--&--&$y$&$x$&$x$&$x^*$&$y^*$&$y^*$&--&--&--\\ \hline
\end{tabular}
}
\end{center}
\end{table*}

\begin{figure}[htb!]
\centering
\includegraphics[width=1.0 \hsize]{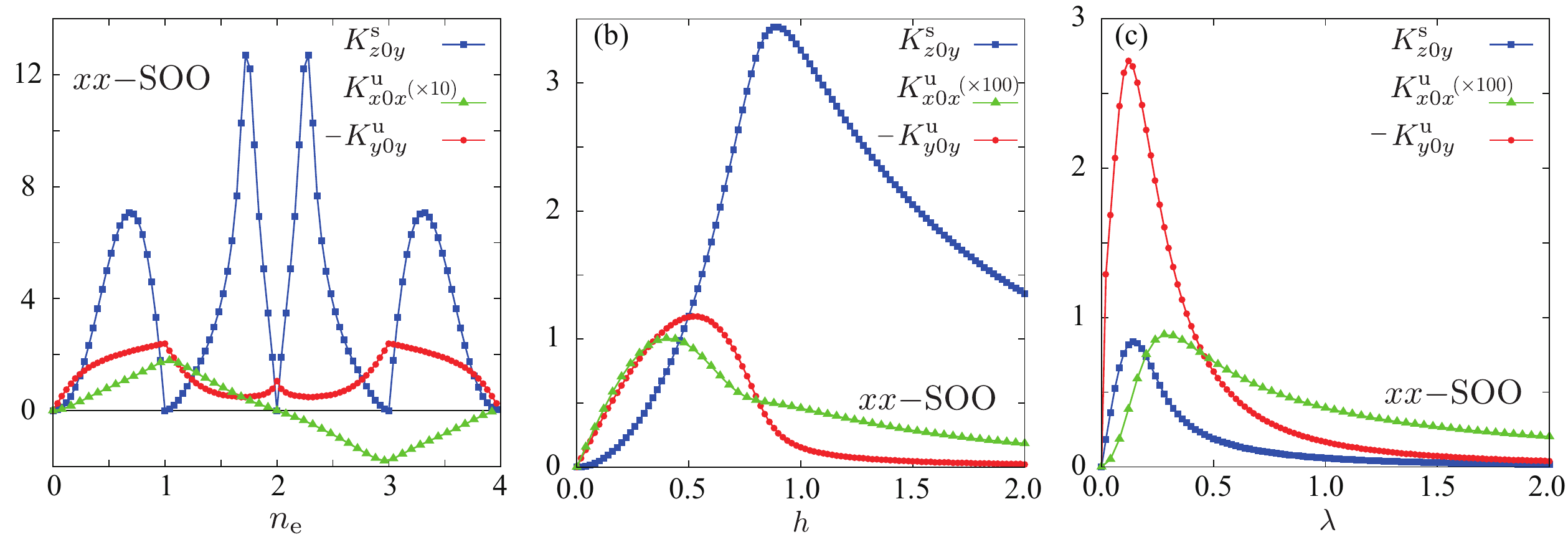} 
\caption{
\label{fig:MEuni}
(a) The coefficients of magnetization-current correlations, $K^{{\rm u}}_{x0x}$, $K^{{\rm u}}_{y0y}$, and $K^{{\rm s}}_{z0y}$, as functions of the electron density $n_{\rm e}$ in the $xx$-SOO (class \#7) phases at $h=0.2$, temperature $T=0.01$, and the damping factor $\delta=0.01$. 
(b) $h$ dependences of $K^{{\rm u}}_{x0x}$, $K^{{\rm u}}_{y0y}$, and $K^{{\rm s}}_{z0y}$ in the $xx$-SOO phase at $n_{{\rm e}}=0.1$.
(c) $\lambda$ dependences of $K^{{\rm u}}_{x0x}$, $K^{{\rm u}}_{y0y}$, and $K^{{\rm s}}_{z0y}$ in the $xx$-SOO phase at $n_{{\rm e}}=0.1$ and $h=0.2$.
}
\end{figure}

Let us first discuss the uniform magnetoelectric responses, i.e., the uniform magnetic moments induced by the electric current, which are described by $K^{{\rm u}}_{\alpha0\mu}$: we consider $m^{nm}_{{\rm u}\alpha0,\bi{k}}$ with $\alpha=x,y,z$ in equation~(\ref{Eq:Mangetoelectric effect_mae}).  
For instance, $K^{{\rm u}}_{x0x}$ is the coefficient for the uniform magnetization in the $x$ direction induced by the electric current in the $x$ direction. 
This is a longitudinal magnetoelectric effect. 
Remarkably, among sixteen possible electronic orders, only the SOO states belonging to the class \#7 exhibit the uniform magnetoelectric effects~\cite{Hayami_PhysRevB.90.081115} [sME(u) in table~\ref{tab}]. 
Each phase in the class \#7 shows two components of $K^{{\rm u}}_{\mu0\nu}$: 
$K^{{\rm u}}_{x0x}$ and $K^{{\rm u}}_{y0y}$ for $xx$- and $yy$-SOO and $K^{{\rm u}}_{x0y}$ and $K^{{\rm u}}_{y0x}$ for $xy$- and $yx$-SOO, respectively.
This result indicates that, in the present honeycomb-lattice model, the uniform magnetoelectric effect is observed when both threefold rotational and mirror symmetries are broken in addition to time-reversal symmetry. 

As a typical example, we show the results for the $xx$-SOO state. 
In this case, $K^{{\rm u}}_{{y0y}}$ becomes nonzero in the entire region of $n_{{\rm e}}$, while $K^{{\rm u}}_{{x0x}}$ becomes zero only for $n_{\rm e} = 2$, as shown in figure~\ref{fig:MEuni}(a). 
The result indicates that a uniform magnetization in the $x$ ($y$) direction can be induced by an electric current in the $x$ ($y$) direction under the $xx$-SOO. 
$K^{{\rm u}}_{{x0x}}$ and $K^{{\rm u}}_{{y0y}}$ become nonzero even in the insulating cases at commensurate fillings $n_{\rm e} = 1$, 2, and 3 (except for $K^{{\rm u}}_{{x0x}}$ at $n_{\rm e}=2$), since the dominant contribution comes from the inter-band components in equation~(\ref{Eq:Mangetoelectric effect_mae}). 
Other phases in the class \#7 also show similar behavior in the different components mentioned above. 
The staggered off-diagonal responses such as $K_{z0y}^{\rm s}$ in figure~\ref{fig:MEuni} will be discussed in the next subsection. 

We also show $h$ and $\lambda$ dependences of the uniform magnetic response to the electric current  at $n_{{\rm e}}=0.1$ in figures~\ref{fig:MEuni}(b) and \ref{fig:MEuni}(c), respectively. 
Both curves show qualitatively similar behavior: 
for small $h$ and $\lambda$, the uniform response increases linearly to $h$ and $\lambda$, while it decreases for large $h$ and $\lambda$ after showing a broad peak. 
The decrease to zero as $h\to\infty$ ($\lambda\to\infty$) is explained by the asymptotic form of the emergent ASOC, which is approximated as $h/(\lambda^2 + h^2) \propto 1/h $ ($\propto h/\lambda^2$) 
[see also equation~(\ref{eq:gxxSOO})].

Next, we discuss uniform responses in the orbital channel. 
The $x$ and $y$ orbital components, $K^{{\rm u}}_{0x\mu}$ and $K^{{\rm u}}_{0y\mu}$, represent the electric quadrupole responses to the electric current. 
We find that the ordered states in the classes \#1, \#2, \#4, and \#5 show nonzero values of $K^{{\rm u}}_{0x\mu}$ and $K^{{\rm u}}_{0y\mu}$. 
Comparing with the emergent ASOC in table~\ref{tab:ASOC}, we note that nonzero responses with $K^{{\rm u}}_{0x\mu}$ and $K^{{\rm u}}_{0y\mu}$ are obtained in the presence of $g^{\rm u}_{\alpha\beta}(\bi{k})$ ($\alpha,\beta=0,z$). 

On the other hand,  
we obtain nonzero $K^{{\rm u}}_{0z\mu}$ in the classes \#2 and \#5 [oME(u) in table~\ref{tab}]. 
The $z$ orbital component, $K^{{\rm u}}_{0z\mu}$, represents the magnetic dipole response, as $\tau_{z}\propto l_{z}$. 
In this case, we find that the breaking of threefold rotational symmetry is  necessary for nonzero $K^{{\rm u}}_{0z\mu}$, in addition to the presence of the emergent ASOC $\propto g^{\rm u}_{\alpha\beta}(\bi{k})$ ($\alpha,\beta=0,z$) (see tables~\ref{tab:ASOC} and \ref{tab:MEuniform}).
Interestingly, there are nonzero orbital-current responses in the ordered states belonging to the class \#2 despite the presence of time-reversal symmetry. 
This is understood by decomposing the electric current operator $J_\mu$ into two parts as 
\begin{eqnarray}
J_{\mu}&=&\frac{\partial \mathcal{H}}{\partial k_\mu} = J_{\mu}^{(0)}+J_{\mu}^{(1)}, 
\end{eqnarray}
where 
\begin{eqnarray}
J_{\mu}^{(0)}&=& -t_0 \sum_{\bi{k}m \sigma} \left(\frac{\partial \gamma_{0,\bi{k}}}{\partial k_{\mu}} 
c^{\dagger}_{{\rm A}\bi{k}m \sigma}c_{{\rm B}\bi{k}m\sigma} +{\rm H.c.}\right), \\
\label{eq:J1}
J_{\mu}^{(1)}&=& -t_1 \sum_{\bi{k}m \sigma} \left(\frac{\partial \gamma_{m,\bi{k}}}{\partial k_{\mu}} 
c^{\dagger}_{{\rm A}\bi{k}m \sigma}c_{{\rm B}\bi{k}\bar{m}\sigma} +{\rm H.c.}\right). 
\end{eqnarray}
The former $J_{\mu}^{(0)}$ represents the current originating from the intra-orbital hopping $t_0$, while the latter $J_{\mu}^{(1)}$ from the inter-orbital $t_1$. 
The latter is the orbital off-diagonal current carrying the orbital angular momentum $\tau_{z}$, which is proportional to $\tau_x$ and $\tau_y$. 
Combining $J_{\mu}^{(1)}$ with the order parameters $\tau_x$ and $\tau_y$ in the class \#2, we obtain a response in the 
$\tau_z$ component, i.e., the nonzero $K^{{\rm u}}_{0z\nu}$. 

In addition to the magnetic and orbital responses, we also have uniform responses in spin-orbital channels. 
All such uniform off-diagonal responses in the paramagnetic and sixteen staggered ordered states are summarized in table~\ref{tab:MEuniform}. 
We note that the responses become nonzero only when $t_1\neq 0$, while some of them remain nonzero even for $\lambda=0$ (indicated by the superscripts $^*$ in table~\ref{tab:MEuniform}). 
In the paramagnetic state, no response is induced by the electric current because there is no uniform ASOC. 
Meanwhile, each ordered phase shows some off-diagonal responses because of the uniform ASOC induced by the spontaneous parity breaking in table~\ref{tab:ASOC}. 
For instance, in the $x$-OO phase in the class \#2, when the electric current is applied in the $x$ direction, the uniform $\langle\tau_{x}\rangle$, $\langle\sigma_{z}\tau_{y}\rangle$, and $\langle\sigma_{z}\tau_{z}\rangle$ moments are induced, while $\langle\tau_{y}\rangle$, $\langle\tau_{z}\rangle$, and $\langle\sigma_{z}\tau_{x}\rangle$ are induced when the electric current is applied in the $y$ direction. 
This indicates that the electric quadrupoles $l_x^2 - l_y^2$ ($l_x l_y + l_y l_x$) are detected in the electric current applied in the $x$ ($y$) direction in the $x$-OO phase. 

\subsection{Staggered response to electric current}
\label{sec:Staggered response by electric current}

\begin{table}[htb!]
\begin{center}
\caption{
Table of staggered off-diagonal responses in paramagnetic and sixteen staggered ordered states. 
The notations are common to table~\ref{tab:MEuniform}. 
}
\label{tab:MEstaggered}
\scalebox{0.76}{
 \begin{tabular}{cccccccccccccccccc}
 \hline
\#& & C & ${\rm S}_x$& ${\rm S}_y$& ${\rm S}_z$  &${\rm O}_x$ &${\rm O}_y$&${\rm O}_z$&${\rm SO}_{xx}$&${\rm SO}_{xy}$&${\rm SO}_{xz}$&${\rm SO}_{yx}$&${\rm SO}_{yy}$&${\rm SO}_{yz}$&${\rm SO}_{zx}$&${\rm SO}_{zy}$&${\rm SO}_{zz}$  \\
\hline\hline
0&PM           &--&--&--&--&$x^*$&$y^*$&--&--&--&--&--&--&--&$y$&$x$&--\\ \hline
\multirow{2}{*}{1}&CO           &--&--&--&--&$x^*$&$y^*$&--&--&--&--&--&--&--&$y$&$x$&--\\ 
&$zz$-SOO&--&--&--&--&$x^*$&$y^*$&--&--&--&--&--&--&--&$y$&$x$&--\\ \hline
\multirow{2}{*}{2}&$x$-OO    &$x^*$&--&--&$y$&$x^*$&$y^*$&$y^*$&--&--&--&--&--&--&$y$&$x$&$x$\\ 
&$y$-OO    &$x^*$&--&--&$y$&$x^*$&$y^*$&$y^*$&--&--&--&--&--&--&$y$&$x$&$x$\\ \hline
\multirow{2}{*}{3}&$xz$-SOO&--&--&--&--&$x^*$&$y^*$&--&--&--&--&--&--&--&$y$&$x$&--\\ 
&$yz$-SOO&--&--&--&--&$x^*$&$y^*$&--&--&--&--&--&--&--&$y$&$x$&--\\ \hline
\multirow{2}{*}{4}&$z$-SO    &--&--&--&--&$x^*$&$y^*$&--&--&--&--&--&--&--&$y$&$x$&--\\ 
&$z$-OO    &--&--&--&--&$x^*$&$y^*$&--&--&--&--&--&--&--&$y$&$x$&--\\ \hline
\multirow{2}{*}{5}&$zx$-SOO&$x^*$&--&--&$y$&$x^*$&$y^*$&$y^*$&--&--&--&--&--&--&$y$&$x$&$x$\\ 
&$zy$-SOO&$x^*$&--&--&$y$&$x^*$&$y^*$&$y^*$&--&--&--&--&--&--&$y$&$x$&$x$\\ \hline
\multirow{2}{*}{6}&$x$-SO    &--&--&--&--&$x^*$&$y^*$&--&--&--&--&--&--&--&$y$&$x$&--\\ 
&$y$-SO    &--&--&--&--&$x^*$&$y^*$&--&--&--&--&--&--&--&$y$&$x$&--\\ \hline
\multirow{4}{*}{7}&$xx$-SOO&$x^*$&--&--&$y$&$x^*$&$y^*$&$y^*$&--&--&--&--&--&--&$y$&$x$&$x$\\ 
&$yy$-SOO&$x^*$&--&--&$y$&$x^*$&$y^*$&$y^*$&--&--&--&--&--&--&$y$&$x$&$x$\\ 
&$xy$-SOO&$x^*$&--&--&$y$&$x^*$&$y^*$&$y^*$&--&--&--&--&--&--&$y$&$x$&$x$\\ 
&$yx$-SOO&$x^*$&--&--&$y$&$x^*$&$y^*$&$y^*$&--&--&--&--&--&--&$y$&$x$&$x$\\ \hline
\end{tabular}
}
\end{center}
\end{table}

\begin{figure}[htb!]
\centering
\includegraphics[width=1.0 \hsize]{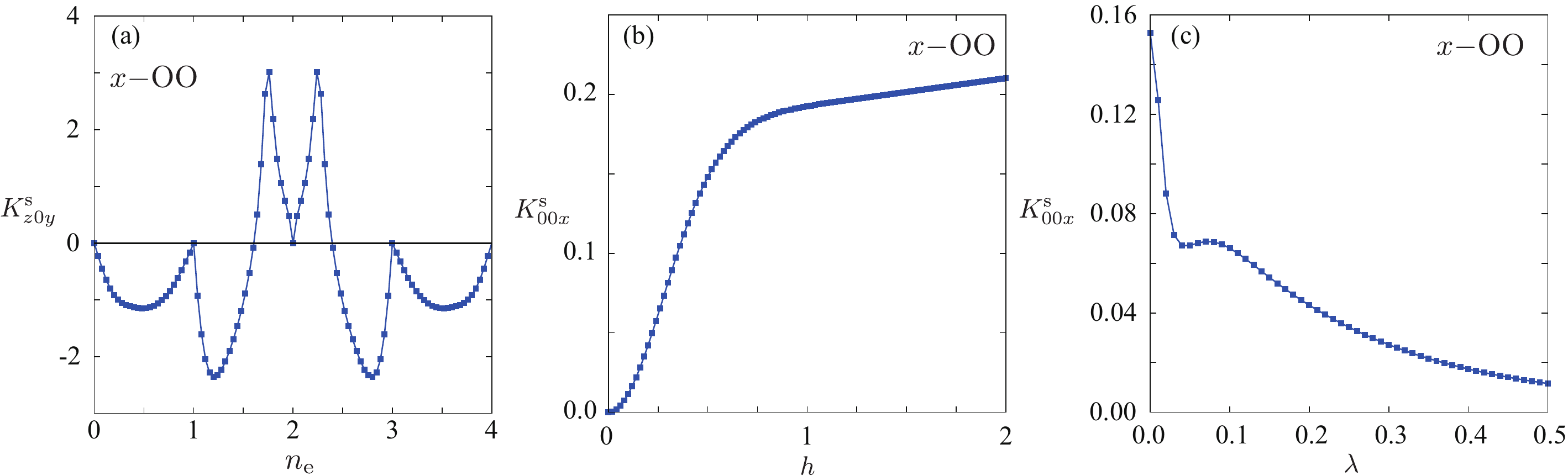} 
\caption{
\label{fig:MEsta}
(a) The coefficient of magnetization-current correlation, $K^{{\rm s}}_{z0y}$, as function of the electron density $n_{\rm e}$ in the $x$-OO and (class \#2) at $h=0.1$, temperature $T=0.01$, and the damping factor $\delta=0.01$. 
(b) $h$ dependence of $K^{{\rm s}}_{00x}$ in the $x$-OO phase at $n_{{\rm e}}=0.1$ and $\lambda=0.5$.
(c) $\lambda$ dependence of $K^{{\rm s}}_{00x}$ in the $x$-OO phase at $n_{{\rm e}}=0.1$ and $h=0.1$.
}
\end{figure}

We turn to the staggered magnetoelectric effects. 
The staggered magnetoelectric responses were discussed for the centrosymmetric systems with local asymmetry, even in the paramagnetic state~\cite{Yanase:JPSJ.83.014703,Hayami_PhysRevB.90.024432}. 
In the present honeycomb-lattice model, however, the linear magnetoelectric effect does not appear as long as the threefold rotational symmetry $\mathcal{R}$ is preserved; once $\mathcal{R}$ is broken by spontaneous electronic ordering, the ASOC is induced with $\bi{k}$-linear contributions (see table~\ref{tab:ASOC}), which gives rise to staggered linear magnetoelectric responses, as indicated by sME(s) in table~\ref{tab}. 
As a typical example, we examine the $x$-OO in the class \#2. 
Figure~\ref{fig:MEsta}(a) shows $K^{{\rm s}}_{z0y}$ as a function of the electron density $n_{{\rm e}}$. 
$K^{{\rm s}}_{z0y}$ becomes nonzero in the entire region of $n_{{\rm e}}$, except when the system is insulating at integer $n_{\rm e}$~\footnote{At $n_{{\rm e}} \sim 1.6$ and $2.4$, $K^{{\rm s}}_{z0y}$ becomes zero in changing the sign although the system is metallic.}. 
The result indicates that the staggered magnetic moment in the $z$ direction is induced by the electric current in the $y$ direction for the $x$-OO order (other staggered magnetic responses are all zero). 
This is because the $x$-OO breaks the rotational symmetry $\mathcal{R}$, which activates 
the ASOC contributing to the linear magnetoelectric term, as discussed in section~\ref{sec:Effect of spontaneous parity breaking}. 

Similar staggered magnetoelectric effects are obtained in the class \#7. 
Figure~\ref{fig:MEuni}(a) shows a staggered response, $K^{{\rm s}}_{z0y}$, as 
a function of $n_{{\rm e}}$. 
In this case also, the staggered magnetic response is induced, consistent with the ASOC in the $xx$-SOO phase in table~\ref{tab:ASOC}.  
We also show $h$ and $\lambda$ dependences of $K^{{\rm s}}_{z0y}$ at $n_{{\rm e}}=0.1$ in figures~\ref{fig:MEuni}(b) and \ref{fig:MEuni}(c), respectively. 
Both curves show similar behavior to the uniform magnetic responses discussed in section~\ref{sec:Uniform response by electric current}: for small $h$ ($\lambda$), the staggered responses increase with increasing $h$ ($\lambda$) due to the emergence of the ASOC, while they gradually decrease and approach to zero for large $h$ ($\lambda$). 
However, in contrast to the uniform ones, the same component $K^{{\rm s}}_{z0y}$ becomes nonzero in the staggered response for all the ordered states in the class \#7. 
Note that the staggered magnetoelectric responses predominantly come from the intra-orbital components in equation~(\ref{Eq:Mangetoelectric effect_mae}), leading to a strong dependence on the damping factor $\delta$. 

Table~\ref{tab:MEstaggered} summarizes the complete table of the staggered off-diagonal responses in the paramagnetic and sixteen staggered ordered states. 
Similar to the uniform ones in table~\ref{tab:MEuniform}, all the nonzero responses appear when $t_1 \neq 0$, while some of them remain nonzero even for $\lambda = 0$, as shown in table~\ref{tab:MEstaggered}. 
As a typical example of the off-diagonal responses except for the magnetoelectric effects, we discuss the staggered CO response in the $x$-OO phase. 
Figures~\ref{fig:MEsta}(b) and \ref{fig:MEsta}(c) show $h$ and $\lambda$ dependences of $K^{\rm s}_{00x}$ at $n_{{\rm e}}=0.1$, respectively. 
The behaviors are different from those in the magnetoelectric effects in figures~\ref{fig:MEuni}(b) and \ref{fig:MEuni}(c). 
In the $h$ dependence in figure~\ref{fig:MEsta}(b), the staggered CO response, $K^{\rm s}_{00x}$, continues to increase as increasing $h$, which indicates that a large response persists in the region where the order parameter in the $x$-OO develops well and almost saturates. 
On the other hand, the $\lambda$ dependence in figure~\ref{fig:MEsta}(c) decreases as $\lambda$ increases. 
The nonzero value of $K^{{\rm s}}_{00x}$ at $\lambda=0$ originates in the effective ASOC induced by the order parameters and inter-orbital hopping $t_1$, which indicates that the atomic SOC $\lambda$ does not play a role in this staggered CO response. 
In fact, the effective ASOC in the $x$-OO phase has the form of $g^{{\rm u}}_{0z}(\bi{k})\rho_{0}\sigma_{0}\tau_{z}\propto (t_0 t_1/h) f_{{\rm E}_1}(\bi{k}) \rho_0 \sigma_0 \tau_z$ in the region $h \gg \lambda$ [cf. equation~(\ref{eq:g0xOO})]. 

\section{Summary and Concluding Remarks}
\label{sec:Summary}

To summarize, we have investigated the effect of electronic orders which break the spatial inversion symmetry spontaneously in the spin-orbital coupled systems on the centrosymmetric lattices with local asymmetry. 
We have clarified how the ASOC is generated by the staggered charge, spin, orbital, and spin-orbital orders, taking a minimal two-orbital model on a honeycomb lattice. 
We derived the explicit form of the effective ASOC for all the cases and analyzed them in detail from the symmetry point of view. 
On the basis of the analysis of the ASOC, we have discussed the nature of each electronic orders as well as the paramagnetic state. 
The results are summarized in table~\ref{tab}. 
In the following, let us briefly review the main results. 

In the paramagnetic state, the ASOC is hidden in the sublattice-dependent form (table~\ref{tab:ASOC_para}). 
The hidden ASOC is activated by spontaneous parity breaking, in a different way depending on the symmetry of the electronic order parameters. 
We classified all the possible staggered orders into the seven classes \#1-\#7 by symmetry (table~\ref{tab}) and derived the effective ASOCs for each case (table~\ref{tab:ASOC}). 
Using the comprehensive table of the ASOC, we have examined the electronic properties, such as the spin and valley splitting of the band structure, and the off-diagonal responses to an external electric current. 
In the classes \#1, \#2, and \#3, we showed that the emergent ASOC gives rise to the antisymmetric spin splitting of the band structure. 
Interestingly, the spin splitting appears in a different manner in different classes, which is understood from the form of the ASOC in each class. 
The topological phase transition was discussed for the CO state in the class \#1. 
We also discussed that the ASOC in the class \#2 leads to peculiar off-diagonal responses in both spin and orbital channels, which originates from the breaking of threefold rotational symmetry. 
On the other hand, in the classes \#4 and \#5, the band structures exhibit peculiar deformation, which is ascribed to the violation of time-reversal symmetry in addition to the spatial inversion symmetry. 
In the class \#4, we showed that the emergent ASOC leads to the valley splitting of the band structure. 
We find a topological transition also in this case, similar to the class \#1.
Meanwhile, in the class \#5, the band bottom shift from the $\Gamma$ point is induced by the $\bi{k}$-linear ASOC, similar to the toroidal ordered cases. 
In this case also, we clarified that the system exhibits the off-diagonal responses in the spin and orbital channels, similar to the class \#2. 
In the class \#7, in which the spin-orbital orders break all the four symmetries considered here, we pointed out that the system exhibits both uniform and staggered magnetoelectric responses, in addition to the staggered orbital response. 
Summarizing the results, we have completed the tables for the uniform and staggered responses of the ordered parameters to an electric current (tables~\ref{tab:MEuniform} and \ref{tab:MEstaggered}). 

Our results provide a comprehensive reference for further exploration of emergent physical properties induced by the spontaneously-generated ASOC. 
Our minimal model includes the essential ingredients for such emergent physics: the atomic SOC, electron hopping between orbitals with different angular momentum, electron correlations, and local asymmetry of the lattice structure. 
The obtained ASOC includes charge, spin, and orbital degrees of freedom, which is the generalization of the conventional ASOC studied in the field of semiconductors and topological insulators. 
In other words, we have extended the ASOC physics to multiband correlated electron systems. 
Our present analysis paves the way for investigating new noncentrosymmetric physics by spontaneous parity breaking, such as new types of electromagnetic and transport properties induced by the emergent ASOC. 

Let us conclude by making several remarks on the future problems. 
One of the intriguing problems is the extension of the present analysis to other degrees of freedom. 
In the present study, we elucidated the effect of the ASOC on the electronic structures and off-diagonal responses by using the spin-charge-orbital coupled model. 
Further peculiar off-diagonal responses can be expected by including the coupling to collective modes, such as magnons from magnetic excitations and phonons from the lattice distortion. 
For example, recent experiments showed that the asymmetric magnon excitations in chiral ferromagnets lead to nonreciprocal magnon propagations~\cite{Iguchi_PhysRevB.92.184419,seki2015nonreciprocal}. 
Although such studies were limited to the noncentrosymmetric systems thus far, similar results will be obtained in a more controllable way for centrosymmetric systems with local asymmetry.
In fact, the authors clarified that the AFM orders on the zigzag chain and honeycomb lattice accompany asymmetric magnon dispersions with respect to the wave vector~\cite{Hayami_doi:10.7566/JPSJ.85.053705}. 
Such extensions may result in controlling the multiferroic off-diagonal phenomena, e.g., the magnetostriction and piezoelectric effect~\cite{wu2014piezoelectricity}. 

Another interesting problem is the physics related with domain and interface in the systems with the spontaneous ASOC~\cite{KhomskiiPhysics.2.20}. 
For instance, in the pyrochlore lattice systems, the surfaces and domains can induce characteristic electronic states and off-diagonal responses~\cite{Arima_doi:10.7566/JPSJ.82.013705,Yamaji_PhysRevX.4.021035}. 
Moreover, a gapless domain state appears on the honeycomb lattice in the presence of the staggered potential~\cite{Yao_PhysRevLett.102.096801}. 
Our results will serve as a good reference for comprehensive understanding of such peculiar surface/domains states. 

Finally, the experimental exploration of the physics of emergent ASOC is an important future problem. 
There are good candidate materials, e.g., trichalcogenides $MX'X_3$ ($M$: transition metal, $X$: chalcogen, $X'=$ P, Si, Ge)~\cite{Ressouche_PhysRevB.82.100408,Sivadas_PhysRevB.91.235425}. 
We have predicted several new phenomena related with the staggered electronic ordering, such as the spin-orbital orders. 
Although our model is a skeleton model and further sophistication is necessary to compare with the experiments, we believe that our analyses dig out the new essential features of the spin-charge-orbital coupled systems, which are possibly explored in monolayer $MXX'_3$.
Moreover, similar interesting physics is expected also in other centrosymmetric lattices with local asymmetry like the zig-zag chain and diamond lattice, which are found in several $f$-electron compounds, such as UGe$_2$~\cite{oikawa1996crystal,saxena2000superconductivity,Huxley_PhysRevB.63.144519}, URhGe~\cite{aoki2001coexistence,levy2005magnetic,Hardy_PhysRevLett.94.247006}, UCoGe~\cite{huy2007superconductivity,Huy_PhysRevLett.100.077002,aoki2009extremely}, $Ln$$M_2$Al$_{10}$ ($Ln=$ Ce, Nd, Gd, Dy, Ho, Er, $M=$ Fe, Ru, Os)~\cite{thiede1998ternary,reehuis2003magnetic,KhalyavinPhysRevB.82.100405,TanidaPhysRevB.84.115128,tanida2010possible,kondo2010magnetization,muro2011magnetic,mignot2011neutron}, $RT_2 X_{20}$ ($R=$ Pr, La, Yb, U, $T=$ Fe, Co, Ti, V, Nb, Ru, Rh, Ir, $X=$ Al, Zn)~\cite{moze1998crystal,Bauer_PhysRevB.74.155118,torikachvili2007six,onimaru2010superconductivity,sakai2011kondo,higashinaka2011single,onimaru2012nonmagnetic,onimaru2012simultaneous,matsubayashi2012pressure,sakai2012superconductivity,Tsujimoto_PhysRevLett.113.267001,ikeura2014anomalous}, and $\beta$-YbAlB$_4$~\cite{macaluso2007crystal,nakatsuji2008superconductivity,matsumoto2011quantum,tomita2015high}. 
Further efforts from both theoretical and experimental sides are highly desired for such exploration.

Our results will also be useful for extending the spin-orbit physics in noncetrosymmetric systems. 
For example, in the monolayer transition metal dichalcogenides $MX_2$ with 2H structure~\cite{jobic1990properties,clarke1978low,ali2014large,qian2014quantum}, the transition metal $M$ and the chalcogen $X$ comprise the honeycomb lattice by aligning in a staggered way. 
Compared to the honeycomb-lattice model in the present study, this corresponds to the CO state. Indeed, the spin splitting in the band structure is observed in the monolayer 2H-$MX_2$~\cite{Xiao_PhysRevLett.99.236809}, as predicted for the class \#1 in our model. 
Then, once, e.g., $z$-SO is realized in the monolayer 2H-$MX_2$, a valley splitting is expected in addition to the spin splitting, as discussed in~\cite{li2013coupling}. 
Such a possibility can be examined by introducing a staggered potential into our model. 
Thus, our results predict the emergent properties by spontaneous electronic orders also in noncentrosymmetric systems in a comprehensive way. 
Similar arguments can be applied to materials with a zincblende-type lattice structure, which can be regarded as the CO state on the diamond lattice.

\ack

This work was supported by JSPS KAKENHI Grant Numbers 24340076, 15K05176, 15H05882 (J-Physics), and 15H05885 (J-Physics), the Strategic Programs for Innovative Research (SPIRE), MEXT, and the Computational Materials Science Initiative (CMSI), Japan.

\appendix

\section{Derivation of Effective Antisymmetric Spin-Orbit Coupling}
\label{Sec:Derivation of Effective Antisymmetric Coupling}
In this appendix, we derive the effective ASOC in the model given by equations~(\ref{Eq:H0}) and (\ref{eq: Ham_MF}). 
To this end, we perform the canonical transformation to obtain the effective ASOC at one of two sublattices by eliminating other sublattice~\cite{landau_quantum,Schrieffer_PhysRev.149.491}.
For this purpose, let us divide our mean-field Hamiltonian into the model space and that connecting to out of the model space as
\begin{eqnarray}
\label{eq:Ham_2mat}
\mathcal{H}_{0}+\tilde{\mathcal{H}}_{1}&=\sum_{\bi{k}ss'\sigma\sigma'mm'}\langle m\sigma|H_{ss'}(\bi{k})|m'\sigma' \rangle c_{s\bi{k}m\sigma}^{\dagger}c_{s'\bi{k}m'\sigma'}^{},
\cr&
H(\bi{k})=
\left(
\begin{array}{cc}
H_{\rm A}(\bi{k}) & 0 \\
0 & H_{\rm B}(\bi{k})
\end{array}
\right) + \left(
\begin{array}{cc}
0 & H'(\bi{k}) \\
H'^{\dagger}(\bi{k}) & 0
\end{array}
\right),
\end{eqnarray}
where the $4\times4$ matrices are given by
 \begin{eqnarray}
&
H_{\rm A,B}(\bi{k})=\frac{\lambda}{2}\sigma_{z}\tau_{z}\mp h\sigma_{\alpha}\tau_{\beta}, \\
&
H'(\bi{k})=-[t_{0}\gamma_{0,\bi{k}}\tau_{0}+\frac{t_{1}}{2}(\gamma_{+1,\bi{k}}\tau_{+}+\gamma_{-1,\bi{k}}\tau_{-})]\sigma_{0}.
\end{eqnarray}
In the lowest order with respect to $H'(\bi{k})$, the effective Hamiltonian for the A sublattice is given by 
\begin{eqnarray}
H_{\rm eff}^{\rm A}(\bi{k})=H_{\rm A} - H'^{}\,H_{\rm B}^{-1}\,H'^{\dagger}. 
\label{eq:effHamA}
\end{eqnarray}
Similarly, the effective Hamiltonian for the B sublattice is given by
\begin{eqnarray}
H_{\rm eff}^{\rm B}(\bi{k})=H_{\rm B} - H'^{\dagger}\,H_{\rm A}^{-1}\,H'^{}. 
\label{eq:effHamB}
\end{eqnarray}

In general, the $8\times8$ effective Hamiltonian $H_{\rm eff}(\bi{k})$ can be expanded in terms of $\sigma_{\alpha}\tau_{\beta}$ in addition to $\rho_{0}$ and $\rho_{z}$ as 
\begin{eqnarray}
\label{eq:effHamASOC}
H_{\rm eff}(\bi{k}) = \sum_{\alpha\beta} 
\left[ 
g^{{\rm u}}_{\alpha\beta} (\bi{k})\rho_0 + g^{{\rm s}}_{\alpha\beta} (\bi{k}) \rho_z
\right] 
\sigma_{\alpha} \tau_{\beta}. 
\end{eqnarray}
Considering equation~(\ref{eq:effHamASOC}) at the A and B sublattices, we have the relations,
\begin{equation}
H_{\rm eff}^{\rm A,B}(\bi{k})=\sum_{\alpha\beta}[g^{\rm u}_{\alpha\beta}(\bi{k})\pm g^{\rm s}_{\alpha\beta}(\bi{k})]\sigma_{\alpha}\tau_{\beta}.
\end{equation}
Solving these relations, we obtain the coefficients in the form
\begin{eqnarray}
\label{eq:effg}
g^{\rm u, s}_{\alpha\beta}(\bi{k}) = \frac{1}{8} {\rm Tr} 
\left[(H^{\rm A}_{\rm eff} \pm H^{\rm B}_{\rm eff})\sigma_{\alpha}\tau_{\beta}\right],
\end{eqnarray}
where $H^{\rm A, B}_{\rm eff}(\bi{k})$ are given by equations~(\ref{eq:effHamA}) and (\ref{eq:effHamB}).

\begin{figure}[htb!]
\centering
\includegraphics[width=1.0 \hsize]{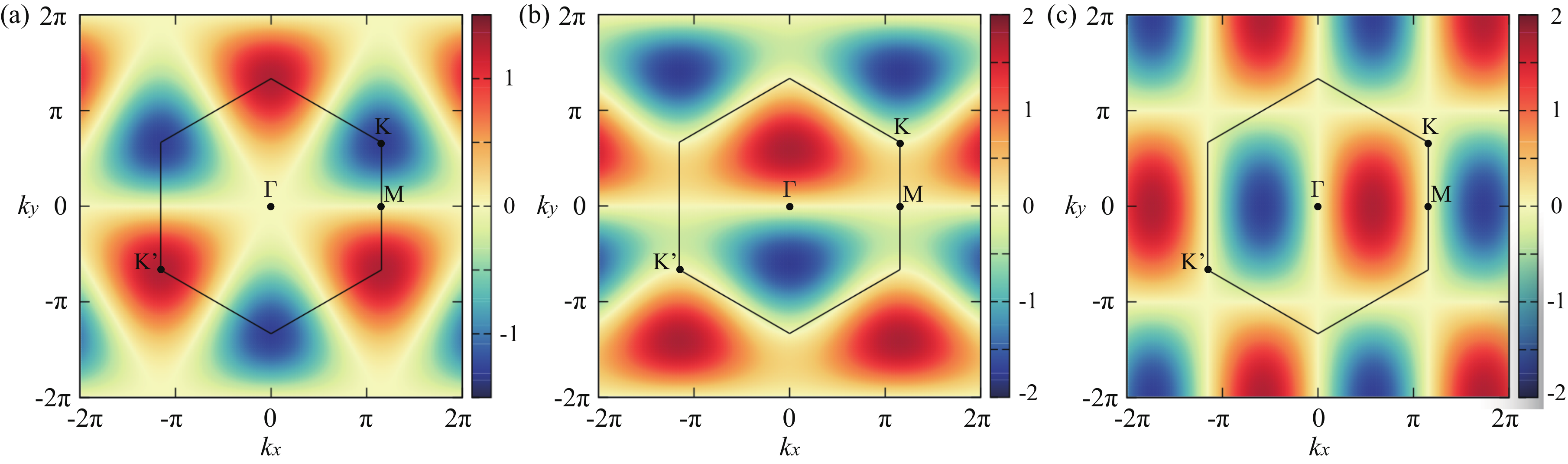} 
\caption{
\label{fig:f0fxfy}
(Color online) 
The contours of (a) $f_{\rm A}(\bi{k})$, (b) $f_{{\rm E}_1}(\bi{k})$, and (c) $f_{{\rm E}_2}(\bi{k})$. 
The hexagons represent the Brillouin zone. 
}
\end{figure}

To demonstrate the usage of equations (\ref{eq:effHamASOC}) and (\ref{eq:effg}), we consider the paramagnetic state ($h=0$).
By the straightforward calculation of equations~(\ref{eq:effHamA}) and (\ref{eq:effHamB}), we have
\begin{eqnarray}
H_{\rm eff}^{{\rm A,B}}=H_{\rm A,B}&\mp\frac{4\sqrt{3}t_{1}^{2}}{\lambda}f_{\rm A}(\bi{k})\sigma_{z}\tau_{0}\mp\frac{4\sqrt{3}t_{0}t_{1}}{\lambda}[f_{{\rm E}_1}(\bi{k})\tau_{x}+f_{{\rm E}_2}(\bi{k})\tau_{y}]\sigma_{z}
\cr&
+\frac{t_{1}^{2}}{\lambda}(|\gamma_{+1,\bi{k}}|^{2}+|\gamma_{-1,\bi{k}}|^{2})\sigma_{z}\tau_{z}-\frac{2t_{0}^{2}}{\lambda}|\gamma_{0,\bi{k}}|^{2}\sigma_{z}\tau_{z},
\end{eqnarray}
where
\begin{eqnarray}
f_{\rm A}(\bi{k})&=-\frac{1}{4\sqrt{3}}(|\gamma_{+1,\bi{k}}|^{2}-|\gamma_{-1,\bi{k}}|^{2})
\cr&
=\left[\cos\left(\frac{\sqrt{3}k_{x}}{2}\right)-\cos\left(\frac{k_{y}}{2}\right)\right]\sin\left(\frac{k_{y}}{2}\right),
\\
f_{{\rm E}_1}(\bi{k})&={\rm Re}f_{\rm E}(\bi{k})
=\left[\cos\left(\frac{\sqrt{3}k_{x}}{2}\right)+2\cos\left(\frac{k_{y}}{2}\right)\right]\sin\left(\frac{k_{y}}{2}\right),
\\
f_{{\rm E}_2}(\bi{k})&=-{\rm Im}f_{\rm E}(\bi{k})
=\sqrt{3}\sin\left(\frac{\sqrt{3}k_{x}}{2}\right)\cos\left(\frac{k_{y}}{2}\right),
\end{eqnarray}
with
\begin{equation}
f_{\rm E}(\bi{k})=-\frac{1}{2\sqrt{3}}(\gamma_{+1,\bi{k}}^{}\gamma_{0,\bi{k}}^{*}-\gamma_{-1,\bi{k}}^{*}\gamma_{0,\bi{k}}^{}).  
\end{equation}
Here, the functions, $f_{\rm A}(\bi{k})$, $f_{{\rm E}_1}(\bi{k})$, and $f_{{\rm E}_2}(\bi{k})$, are all antisymmetric with respect to $\bi{k}$, as shown in figure~\ref{fig:f0fxfy}.
Thus, the uniform components of the effective Hamiltonian are symmetric with respect to $\bi{k}$.
The staggered components are antisymmetric and they are given by
\begin{equation}
H_{\rm eff}^{\rm ASOC}=
-\frac{4\sqrt{3}t_{1}^{2}}{\lambda}f_{\rm A}(\bi{k})\sigma_{z}\tau_{0}-\frac{4\sqrt{3}t_{0}t_{1}}{\lambda}[f_{{\rm E}_1}(\bi{k})\tau_{x}+f_{{\rm E}_2}(\bi{k})\tau_{y}]\sigma_{z}.
\end{equation}

\section{Eigenvalue analysis}
\label{sec:Comparison between the antisymmetric spin-orbit coupling and eigenvalues}

\begin{figure}[htb!]
\centering
\includegraphics[width=0.75 \hsize]{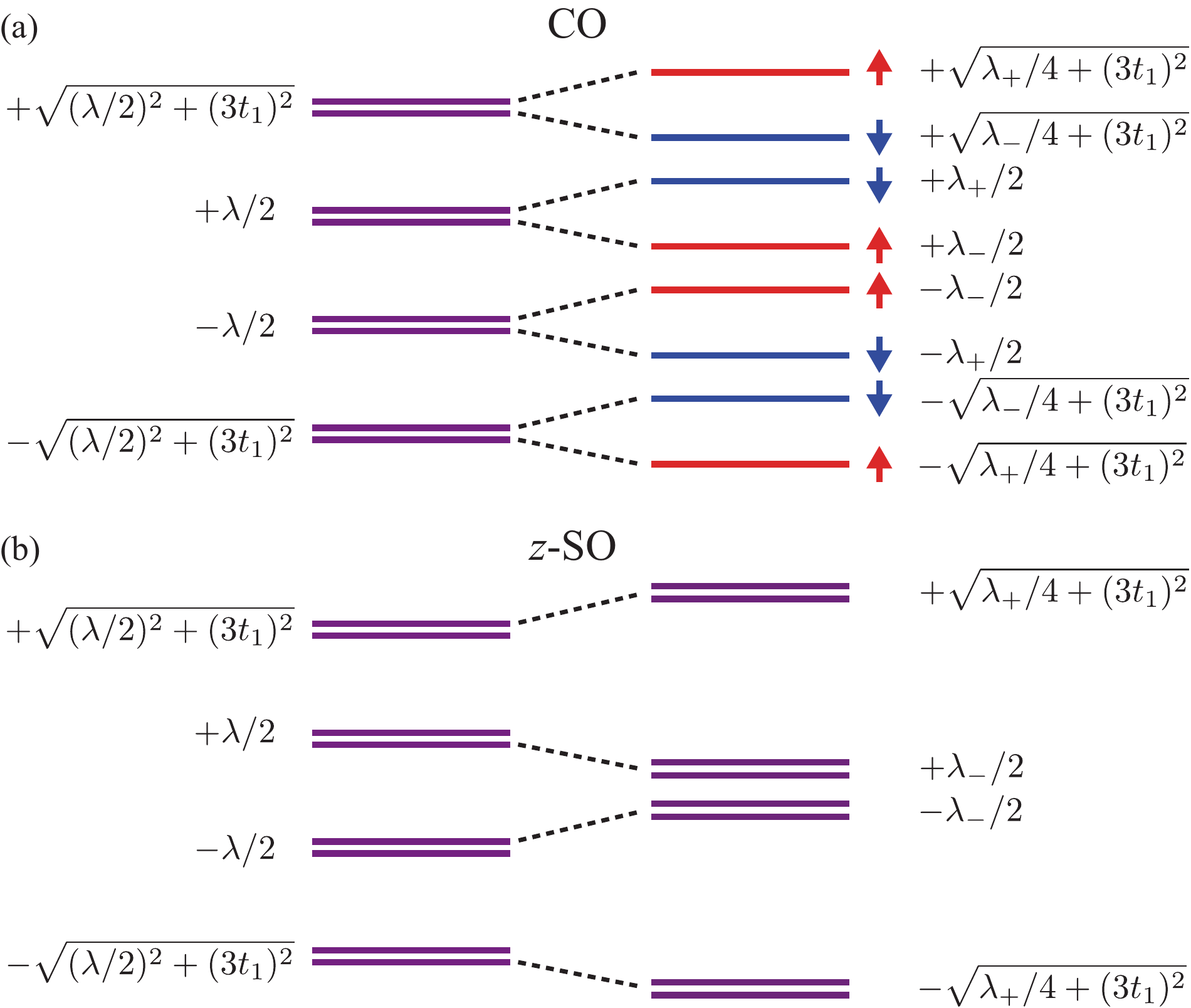} 
\caption{
\label{fig:energylevel_CO}
Schematic diagram of the energy eigenvalues at the K' point when introducing (a) CO (class \#1) and (b) $z$-SO (class \#4). The red (blue) levels show the bands with up(down)-spin polarization. 
}
\end{figure}

In this appendix, we reexamine the effect of ASOC from the viewpoint of eigenvalues of the Hamiltonian.
We discuss the correspondence between the ASOC derived from equation~(\ref{eq:effg}) and the energy eigenvalues by direct diagonalization of the mean-field Hamiltonian, $\mathcal{H}_0 + \tilde{\mathcal{H}}_1$. 
We show that the peculiar electronic structures, such as the antisymmetric spin splitting and band deformation, are understood from the eigenvalue analysis. 

First, we consider the CO state in section~\ref{sec:Class1} (class \#1). 
The antisymmetric spin splitting is derived by directly calculating the energy eigenvalues at the K and K' points in the CO state. 
At the K' point, the eigenvalues are easily obtained by the diagonalization of $\mathcal{H}_0 + \tilde{\mathcal{H}}_1$ as 
\begin{eqnarray}
\label{eq:CO_level}
\varepsilon^{\uparrow}_{{\rm CO}} ({\rm K'}) = &\pm \sqrt{\frac{\lambda_+^2}{4}+(3t_1)^2}, \ \pm \frac{\lambda_-}{2},
\cr
\varepsilon^{\downarrow}_{{\rm CO}} ({\rm K'}) = &\pm \sqrt{\frac{\lambda_-^2}{4}+(3t_1)^2}, \ \pm \frac{\lambda_+}{2},
\end{eqnarray}
where $\lambda_{\pm}=\lambda \pm 2 h$. 
The energy levels are schematically displayed in figure~\ref{fig:energylevel_CO}(a). 
The eigenvalues clearly show that the spin splitting takes place under the CO at the K' point, which is consistent with the argument related to the effective ASOC in equation~(\ref{eq:gCO}). 
The opposite spin splitting occurs at the K point because $h$ changes into $-h$ at the K point. 

Similarly, the valley splitting in the $z$-SO state discussed in section~\ref{sec:Valley splitting} (class \#4) is understood from the energy eigenvalues at the K and K' points. 
The eigenvalues in the $z$-SO case are obtained by changing the sign of the $h$ terms for down spins in those for the CO case above, namely, 
\begin{eqnarray}
\varepsilon_{z{\rm 
-SO}}^{\sigma}({\rm K'}) &= \pm \sqrt{\frac{\lambda_+^2}{4} + (3t_1)^2}, \ \pm \frac{\lambda_-}{2}, \\
\varepsilon_{z{\rm 
-SO}}^{\sigma}({\rm K}) &= \pm \sqrt{\frac{\lambda_-^2}{4} + (3t_1)^2}, \ \pm \frac{\lambda_+}{2}, 
\end{eqnarray}
where $\sigma$ represents +1 (-1) for up (down) spin.
Thus, each eigenvalue is doubly degenerate in terms of spin, as schematically shown in figure~\ref{fig:energylevel_CO}(b). 
The result explains the valley splitting; for example, at half filling, the gap at the K' point is $\lambda-2h$, while that at the K point is $\lambda+2h$. 

Next, we discuss the $x$-OO phase (class \#2). 
The $k$-linear contribution in the emergent ASOC in this state discussed in section~\ref{sec:Class2} is also explained by directly examining the eigenvalues of the Hamiltonian. 
For the $x$-OO phase ($\tilde{\mathcal{H}}_1 \propto \rho_z \sigma_0 \tau_x$), we can expand the eigenvalues with respect to $h$ and $t_1$ up to the first order: 
\begin{eqnarray}
\epsilon_{x{\rm -OO}}^{\sigma}(\bi{k}) = & \pm\left(D_{\bi{k}}^{-}-\sigma \frac{\sqrt{3}ht_1}{D_{\bi{k}}^{-}|\gamma_{0,\bi{k}}|} f_{{\rm E}_1}(\bi{k})\right),
\cr&
\pm\left(D_{\bi{k}}^{+}+\sigma \frac{\sqrt{3}ht_1}{D_{\bi{k}}^{+}|\gamma_{0,\bi{k}}|} f_{{\rm E}_1}(\bi{k})\right),
\end{eqnarray}
where $\sigma$ represents +1 (-1) for up (down) spin and $D_{\bi{k}\pm}$ is defined by 
\begin{eqnarray}
D_{\bi{k}}^{\pm}= \left|\frac{\lambda}{2}\pm t_{0}|\gamma_{0,\bi{k}}|\right|.
\label{eq:D_pmk}
\end{eqnarray}
Therefore, the change of the eigenvalues by introducing $h$ is represented by 
\begin{eqnarray}
\Delta\epsilon_{x-{\rm OO}}^{\sigma}(\bi{k}) \sim \sigma ht_{1} f_{{\rm E}_1}(\bi{k}). 
\end{eqnarray}
In the limit of $\bi{k}\rightarrow 0$, we obtain 
\begin{eqnarray}
\label{eq:H-ASOC-xOO}
\Delta\epsilon_{x-{\rm OO}}^{\sigma}(\bi{k}) \sim \sigma ht_{1} k_y. 
\end{eqnarray}
The result in equation~(\ref{eq:H-ASOC-xOO}) gives the consistent dependence on $h$, $t_1$, and $k_y$ with equation~(\ref{eq:g0xOO}). 
Similarly, the eigenvalues in the limit of $\bi{k}\rightarrow 0$ for the $y$-OO state is obtained in the consistent form with equation~(\ref{eq:g0yOO}) as
\begin{eqnarray}
\label{eq:H-ASOC-yOO}
\Delta\varepsilon_{y-{\rm OO}}^{\sigma}(\bi{k}) \sim \sigma ht_{1} k_x.   
\end{eqnarray}

Now, we turn to the class \#5. 
The effective ASOC in this case is derived from the diagonalization by using the fact that the Hamiltonian in the $zx$-SOO phase has a similar form to that in the $x$-OO phase in the class \#2, with a difference in the presence of $\sigma_z$ in $\tilde{\mathcal{H}}_1$. 
Hence, the eigenvalues in the $zx$-SOO state are obtained by replacing $h$ with $\sigma h$ in the results of the class \#2. 
Then, the change of the eigenvalues by the $zx$-SOO in the limit of $\bi{k}\rightarrow 0$ reads 
\begin{eqnarray}
\label{eq:H-ASOC-zxSOO}
\Delta\varepsilon_{zx-{\rm SOO}}^{\sigma}(\bi{k}) \sim h t_1 k_y.  
\end{eqnarray}
Similar to the class \#2, the eigenvalues are proportional to $k_y$, but without any spin dependence. 
This explains the asymmetric band modulation with a band bottom shift in the $k_y$ direction discussed in section~\ref{sec:Band deformation}.

So far, we have shown that the spin splitting, valley splitting, and the band deformation in the band structure are well understood by performing the direct diagonalization. 
In some cases, however, the eigenvalue analysis does not work out properly. 
We here consider the band structure in the $xz$-SOO case in section~\ref{sec:Class3} (class \#3). 
By the similar procedure of the previous examples, we can derive the asymptotic form of the eigenvalue in the limit of $\bi{k} \rightarrow 0$ by diagonalizing the Hamiltonian: 
\begin{eqnarray}
\label{eq:H-ASOC-xzSOO}
\Delta\epsilon_{xz-{\rm SOO}}^{\sigma}(\bi{k}) \propto  h t_1 \sqrt{k_x^2+k_y^2}.
\end{eqnarray}
The isotropic $\bi{k}$ dependence in equation~(\ref{eq:H-ASOC-xzSOO}) indicates that the band structures are modulated in the symmetric way in this case.  
Hence, we can extract no information about the ASOC in the $xz$-SOO state, in contrast to the discussion given in section~\ref{sec:Class3}. 
Furthermore, any band modulations do not seem to occur by equation~(\ref{eq:H-ASOC-xzSOO}), although the $xz$-SOO shows the spin splitting, as shown in figure~\ref{fig:band_splitting}(h). 
This example signals the failure of the eigenvalue analysis. 
Indeed, the eigenvalues of the equation~(\ref{eq:g0class3}) are isotropic, and only the eigenvectors have the information about their spin dependences.
Meanwhile, the derivation of the effective ASOC by equation~(\ref{eq:effg}) is straightforward and gives comprehensive understanding of the modulations of the band structure and the off-diagonal responses to an electric current.

\bibliographystyle{iopart-num}
\bibliography{ref}

\end{document}